\documentclass[aps, twocolumn,
reprint,
superscriptaddress,
nofootinbib,
amsmath,amssymb,
aps,
pra,
floats,floatfix,english]{revtex4-2}

\usepackage{graphicx}
\usepackage{dcolumn}
\usepackage{bm}

\usepackage[T1]{fontenc}
\usepackage[utf8]{inputenc}
\usepackage[dvipsnames,svgnames,x11names,hyperref]{xcolor}
\usepackage{bm}
\usepackage{booktabs}
\usepackage{graphicx}
\usepackage{mathtools}
\usepackage{mhchem}[version=newest]
\usepackage{multirow}

\usepackage{bbm}
\usepackage{dsfont}
\usepackage[caption=false]{subfig}
\usepackage{tikz}
\usetikzlibrary{arrows,arrows.meta}
\usepackage[singlespacing]{setspace}
\usepackage[squaren]{SIunits}
\usepackage{wasysym} 
\usepackage{hyperref}
\hypersetup{colorlinks=true,linkcolor=blue,citecolor=blue,urlcolor=blue}

\begin{document}

\preprint{APS/123-QED}

\title{Localization landscape for interacting Bose gases\\ in one-dimensional speckle potentials}

\author{Filippo Stellin}
\email{filippo.stellin@ens-paris-saclay.fr}

\affiliation{%
Université Paris-Saclay, CNRS, ENS Paris-Saclay, Centre Borelli, 91190, Gif-sur-Yvette, France.
}
\author{Marcel Filoche}%
\email{marcel.filoche@espci.psl.eu}

\affiliation{Institut Langevin, ESPCI Paris, Université PSL, CNRS, 75005 Paris, France.}
\affiliation{Laboratoire de Physique de la Matière Condensée, École Polytechnique,\\ CNRS, Institut Polytechnique de Paris, Palaiseau, 91120, France.
}
\author{Frédéric Dias}
\email{frederic.dias@ucd.ie}

\affiliation{%
Université Paris-Saclay, CNRS, ENS Paris-Saclay, Centre Borelli, 91190, Gif-sur-Yvette, France.
}%
\affiliation{School of Mathematics and Statistics, University College Dublin, Belfield, Dublin 4, Ireland.}


\date{\today}

\begin{abstract}

While the properties and the shape of the ground state of a gas of ultracold bosons are well understood in harmonic potentials, they remain for a large part unknown in the case of random potentials. Here, we use the localization-landscape (LL) theory to study the properties of the solutions to the Gross-Pitaevskii equation (GPE) in one-dimensional (1D) speckle potentials. In the cases of attractive interactions, we find that the LL allows one to predict the position of the localization center of the ground state (GS) of the GPE. For weakly repulsive interactions, we point out that the GS of the quasi-1D GPE can be understood as a superposition of a finite number of single-particle states, which can be computed by exploiting the LL. For intermediate repulsive interactions, we introduce a Thomas-Fermi-like approach for the GS which holds in the smoothing regime, well beyond the usual approximation involving the original potential.
 Moreover, we show that, in the Lifshitz glass regime, the particle density and the chemical potential can be well estimated by the LL. Our approach can be applied to any positive-valued random potential endowed with finite-range correlations and can be generalized to higher-dimensional systems.

\end{abstract}

\maketitle

\section{\label{sec:intro}Introduction}

Cold atom experiments are a remarkable platform to explore quantum theories and to test open questions in condensed matter physics. The modern developments in cooling and trapping techniques~\cite{LewSanAhu:6:2007} have enabled to achieve Bose-Einstein condensation of matter waves~\cite{AndersonWiemanCornell:BEC:1995, Stringari:BECRevTh:RMP99}, thus opening the possibility to study their behavior in random optical potentials~\cite{Billy:AndersonBEC1D:N08,Roati:AubryAndreBEC1D:N08,Jendrzejewski:AndersonLoc3D:NP12,VolPasDen:8:2018}. Bose-Einstein condensates, occurring in dilute samples and at very low temperature, of the order of tens of $\mathrm{nK}$, are characterized by the macroscopic occupation of the ground state (GS) of the gas, described by a highly coherent and fully symmetric wavefunction. In the absence of interactions among atoms, the interference between the multiple scattering paths of an initially traveling particle can completely inhibit its diffusion, eventually leading to an exponential localization of the wavefunction~\cite{Anderson:1958, Evers:AndersonTransitions:RMP08,Abrahams:2010}. This phenomenon, known as \emph{Anderson localization}, has been actually observed~\cite{Roati:AubryAndreBEC1D:N08,Billy:AndersonBEC1D:N08,Jendrzejewski:AndersonLoc3D:NP12,Kondov:PRL2015} and theoretically studied~\cite{Shapiro:JPA:2012,Piraud:MobilityEdge3D:NJP13,Delande:MobEdgeSpeckle:PRL2014,LSPAspect:BECALEstimates:2007} with matter waves in different settings during the past 15 years.

The presence of interactions between atoms can significantly modify this picture. The study of the interplay between an external quenched disorder and an interacting Bose gas has motivated many theoretical~\cite{GiamSchu:1988, FisWeiGr:4:1989, GeigerBuchleitnerWellens:2013, KhelPels:2016, KhelPels:2017} and experimental works~\cite{LyeFalMod:6:2005, SchDreKru:8:2005, ClementAspect:Quasi1DBEC:PRA2007, CheHitDri:6:2008, DeiZacRoa:8:2010, Guillamon:VortexStability2DDis:2014, BoiBerFou:7:2017}. The many-body interactions make computations of the many-particle states incredibly much harder. However, by treating the interactions through a mean-field approach, one can reduce the dimensionality of the problem and model the gas by a one-particle nonlinear Schrödinger equation, also called the \emph{Gross-Pitaevskii equation} (GPE)~\cite{SanchezPalencia:DisorderQGases:NP10}. Theoretical investigations on these systems have been carried out from different perspectives, focusing on stationary states~\cite{SanchezPalencia:Disorder1DGPEThomasFermi:2006, NattPokr:2008,AkkGhoMus:2008,FalNatPok:2009, ChengAdhikari:GPEstatsolnum:2010}, excitations~\cite{BilasPavloff:2006, GurRefCha:2008, FonWouSav:2009, GauRenMul:2009}, dynamics~\cite{Modugno:dynamicsDisBEC:2006, PalpSucc:2008,JoshGhos:2009, DonHofKoc:5:2017, Najafabadi:TransportBECDis:2021}, out-of-equilibrium physics~\cite{Scoquart:WeakDisGPE:2020, ScoquartBis:GPEDis:2020, cherroretScoquart_Scattering:2021}, phase transitions~\cite{LugCleBou:6:2007, AstKruNav:2013, CarBoeHol:4:2013, Saliba:SFBoseGlassDis2D:2014}, superfluidity~\cite{AlbeMull:2020} and solitons~\cite{Delande:BirghtSolitonLoc1D:2009, PlodSach:2012, Sacha:DarkSolitonDis:2012}.

While the stationary states of the GPE in the weakly interacting limit~\cite{ChengAdhikari:GPEstatsolnum:2010,LugCleBou:6:2007} and the Thomas-Fermi limit for repulsive interactions~\cite{SanchezPalencia:Disorder1DGPEThomasFermi:2006,ChengAdhikari:GPEstatsolnum:2010} are quite well understood, we still lack to this day theoretical tools to tackle the intermediate regime and the strongly attractive limit. 

In the case of strong repulsive interactions, it was shown that, for a chemical potential $\mu$ much larger than the standard deviation of the disorder $V_{0}$, the kinetic term of the GPE can be neglected according to the Thomas-Fermi approximation~\cite{Stringari:BECRevTh:RMP99,AftDalJos:2012}.
Hence, the macroscopic state at equilibrium follows the modulations of the random potential, as it was shown by considering speckle potentials~\cite{SanchezPalencia:Disorder1DGPEThomasFermi:2006, AstKruNav:2013,AkkGhoMus:2008} and Gaussian random potentials~\cite{KhelPels:2016}. In correlated speckle potentials, Sanchez-Palencia pointed out that the stationary state becomes sensitive to a smoothed random potential~\cite{SanchezPalencia:Disorder1DGPEThomasFermi:2006} rather than the original one. This was predicted to take place when the correlation length $\sigma$ is smaller than the healing length $\xi$, i.e., the maximum length of the spatial variations of the state $\psi$ that contribute to the kinetic energy of the atoms~\cite{Stringari:BECRevTh:RMP99}.
In the presence of disorder, the integrated density of states (IDoS), which is the number of states below a given energy, exhibits a low-energy drop characterized by a stretched exponential behavior, known as Lifshitz tail~\cite{Lifshitz:IDoSTails:1963,LifGrePas:1988}, which is related to the existence of large regions of negligible potential~\cite{FalFedGia:4:2010}. In 2007, Lugan \textit{et al.}~\cite{LugCleBou:6:2007} proposed a schematic quantum-state diagram and predicted in the same geometry the occurrence of the Lifshitz glass phase for strong disorder and weakly repulsive interactions. In this regime, the Bose gas splits into fragments whose shapes are given by nonoverlapping single-particle (SP) states belonging to the Lifshitz tails of the IDoS.

Regime diagrams for the repulsive case were also built by Falco \textit{et al.}~\cite{NattPokr:2008, FalNatPok:2009}. The authors examined random potentials with unbounded probability distributions and different correlation profiles, superimposed with harmonic potentials along all directions. Four different regimes were distinguished, based on the spatial behavior of the particle density: harmonic, Thomas-Fermi, nonergodic and fragmented regimes.
However, in the aforementioned studies the collective $N$-particle states were not computed explicitly, but estimates of the typical size of the atomic cloud or of its fragments were provided as functions of the characteristic lengths associated to the random potential and of the coefficient of the nonlinear term $g(N-1)$ of the GPE.

A numerical investigation of the ground state $\psi_{0}$ for weakly repulsive and attractive interactions was carried out in the work of Cheng~\textit{et al.}~\cite{ChengAdhikari:GPEstatsolnum:2010}, who considered 1D~speckle potentials and showed that, for weak interactions, the state remains exponentially localized with a localization length that increases for stationary solutions of the defocusing GPE, whereas it decreases for the focusing GPE.

The work presented here intends to fill the gap of knowledge between the noninteracting and the Thomas-Fermi regime as well as in the strongly attractive limit by exploiting the concept of localization landscape (LL)~\cite{FilocheMayboroda:PNAS2012} which was initially introduced for the non-interacting problem. We exhibit analytical approximations of the many-particle state in 1D speckle potentials in several cases, and also unveil connections with the SP states which were previously analyzed merely in fully harmonic potentials~\cite{Agosta_GPESols_LinSschrDW:2000, Kivshar:PertGPE:2001} and in symmetric double-well potentials~\cite{Agosta_GPESols_NonLinDW:2000}.

\begin{figure}
\centering 
  
    \begin{tikzpicture}
\draw[opacity=0.333,fill=SkyBlue,color=SkyBlue] (4.0,0.0)--(8.0,0.0)--(8.0,4.0)--(4.0,4.0);   
\draw[opacity=0.333,fill=Apricot,color=Apricot] (0.0,0.0)--(4.0,0.0)--(4.0,4.0)--(0.0,4.0);     
\draw[line width=2.0mm, thick, ->,-{Stealth[length=3mm, width=2mm]}] (0.0,0.0) -- (8.0,0.0);
\draw[line width=2.0mm, thick, ->,-{Stealth[length=3mm, width=2mm]}] (4.0,0.0) -- (4.0,4.0);
\node at (4.0,-0.25) {\small{0}};
\node at (4.75,0.5) {\hyperref[subsubsec:wearepint]{\textcolor{black}{IV\hspace{0.025cm}A\hspace{0.025cm}1}}}; 
\node at (7.25,0.5) {\hyperref[subsubsec:strrepint]{\textcolor{black}{IV\hspace{0.025cm}A\hspace{0.025cm}2}}}; 
\node at (5.25,3.25) {\hyperref[subsec:repvardis]{\textcolor{black}{IV\hspace{0.025cm}B}}}; 
\node at (6.0,2.15) {\hyperref[sec:repint]{\textcolor{RoyalBlue}{\small{repulsive}}}};
\node at (6.0,1.85) {\hyperref[sec:repint]{\textcolor{RoyalBlue}{\small{interactions}}}};
\node at (3.0,2.75) {\hyperref[subsec:attvardis]{\textcolor{black}{III\hspace{0.025cm}B}}}; 
\node at (1.0,0.75) {\hyperref[subsec:attvarint]{\textcolor{black}{III\hspace{0.025cm}A}}}; 
\node at (2.0,2.15) {\hyperref[sec:attint]{\textcolor{RedOrange}{\small{attractive}} } };
\node at (2.0,1.85) {\hyperref[sec:attint]{\textcolor{RedOrange}{\small{interactions}} } };
\node at (7.375,-0.375) {\small{$g(N-1)$}};
\node at (4.0,4.25) {\small{$V_{0}$}};  
\end{tikzpicture}
    \caption{Interaction-disorder diagram in which the numbers of the sections pinpoint the regions of the plane where the ground state of the GPE is examined.}
\label{fig:Disorder-Interactions_ArticleStruct}
\end{figure}
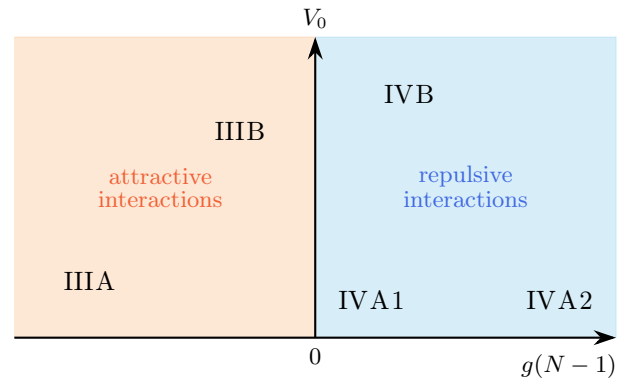

Our paper is structured as follows. In Sec.~\ref{sec:modemeth} we present the physical system, introducing the Gross-Pitaevskii equation, the features of the correlated random potential and the LL function used throughout the work. In the remaining sections, we investigate the regions in the interaction-disorder plane depicted in Fig.~\ref{fig:Disorder-Interactions_ArticleStruct}.

In Sec.~\ref{sec:attint}, we examine the attractive case and we point out that the localization center of the GS of the GPE is well predicted by the absolute minimum of the effective potential. We also numerically show that the disorder-averaged localization length decreases as the nonlinear coefficient or the disorder parameter are increased.

Sec.~\ref{sec:repint} is devoted to the case of repulsive interactions. Here the ground state of the GPE is displayed for a wide range of random potential amplitudes and nonlinear coefficients. In Sec.~\ref{subsubsec:wearepint} and in Appendices~\hyperref[app:coeffeval]{A}-\hyperref[app:loclandSPstates]{B}, we illustrate that the GS of the GPE for weakly repulsive interactions and speckle potentials can be predicted by an expansion over a finite number of (localized) SP states. The quality of the LL approach is assessed by comparing those states, computed by exact diagonalization of the SP Hamiltonian, against the solutions of the eigenvalue problem restricted to the regions of the lowest minima of the effective potential derived from the LL~\cite{Arnold:EffConfLL:2016}. 

In Sec.~\ref{subsubsec:strrepint}, we show that, for intermediate repulsive interactions, the LL-based effective potential sets the typical variation scale of the macroscopic wavefunction. We point out that the last mentioned quantity can be well approximated using a Thomas-Fermi-like ansatz. 

In Sec.~\ref{subsec:repvardis}, we compute the GS in a regime where it is given by a superposition of SP wavefunctions which do not spatially overlap with each other and whose energy lies in the Lifshitz tails of the IDoS. In this case, we show that the number of particles in each of the SP wavefunctions, occurring at the wells of the effective potential, are well predicted by the LL starting from a relation derived by Lugan \textit{et al.}~\cite{LugCleBou:6:2007}.

In Sec.~\ref{sec:conclupers}, we draw the conclusions and outline the possible perspectives of this work.\\

\section{The localization landscape of the Gross-Pitaevskii equation}
\label{sec:modemeth}

\subsection{The Gross-Pitaevskii equation}
\label{subsec:GPE}

We consider an ultracold dilute Bose gas in which interaction events are \emph{binary}, i.e., involve only two particles at a time, and are characterized by a length scale that is smaller than the de Broglie wavelength so that the scattering events are dominated by $s$-wave processes.
Under these conditions, the ground state of the many-particle system is given by the common wavefunction $\psi(\pmb{r})$, normalized to unity, which obeys the Gross-Pitaevskii equation: 
\begin{multline}
\label{eqn:001}
\biggl[-\frac{\hbar^{2}}{2m}\nabla^{2}+V(\pmb{r})\biggr]\psi(\pmb{r})+\\+\frac{4\pi\hbar^{2}a_{s}}{m}(N-1)|\psi (\pmb{r})|^{2}\psi (\pmb{r})=\mu\psi (\pmb{r})\hspace{0.1cm}\text{,}
\end{multline}
where $a_{s}$ represents the $s$-wave scattering length. To ensure the validity of Eq.~\eqref{eqn:001} for a quantum gas, $a_{s}$ must satisfy a low-density assumption, $\langle\rho\rangle |a_{s}|^{3}\ll 1$~\cite{Stringari:BECRevTh:RMP99}, $\langle \rho\rangle=N\langle|\psi|^{2} \rangle$ being the spatial average of the particle density. Moreover, one has to note that this nonlinear coupling can be experimentally tuned by means of Feshbach resonances~\cite{FalForIng:2008}. 
By applying a tight harmonic confinement on one or two dimensions~\cite{SanchezPalencia:Disorder1DGPEThomasFermi:2006, LugCleBou:6:2007}, it is possible to assess the effect of disorder on the macroscopic wavefunction in low-dimensional quantum gases. In the following, we focus on potentials exhibiting a 1D disorder along $Ox$, which means that the total potential can be written as
\begin{equation}
\label{eqn:002}
V_{\rm{tot}}(x,y,z) = V_{\mathrm{R}}(x) + \frac{1}{2} m \omega_{\perp}^2 \left( y^2 + z^2 \right) \,,
\end{equation}
where $V_{\mathrm{R}}$ is the 1D random potential and $\omega_{\perp}$ is the pulsation of the two-dimensional (2D) harmonic well in the directions $y$ and $z$. Assuming that $\hbar\omega_{\perp}$ is much larger than the level spacing between two consecutive eigenvalues of the noninteracting 1D problem, the ground state of this potential can be factorized as
\begin{equation}
\label{eqn:6574}
 \psi_{0}(\pmb{r}) = \psi_{0}(x) \, \phi_{0}^{(\mathrm{ho})}(y,z) \, ,
\end{equation}
where 
\begin{equation}
\label{eqn:060}
\phi_{0}^{(\mathrm{ho})}(y,z)=\sqrt{\frac{m\omega_{\perp}}{\pi \hbar}}\exp{\biggl\{-\frac{m\omega_{\perp}}{2\hbar}(y^{2}+z^{2})\biggr\}}\hspace{0.05cm}\text{,}
\end{equation}
is the ground state of the 2D harmonic oscillator in the $(y,z)$ plane. The two wavefunctions $\psi_{0}$ and $\phi_{0}^{(\mathrm{ho})}$ satisfy the normalization conditions $\int \mathrm{d}x\hspace{0.05cm} |\psi_{0}(x)|^{2}=1$ and $\int \mathrm{d}y\mathrm{d}z\hspace{0.05cm}|\phi_{0}^{(\mathrm{ho})}(y,z)|^{2}=1$, respectively. After integrating out the 2D harmonic wavefunction, one finds that $\psi_{0}$ obeys~\cite{SanchezPalencia:Disorder1DGPEThomasFermi:2006,ZamoraParedes:GPE_HarmonicNumSol:2019}
\begin{multline}
\label{eqn:054}
\biggl[-\frac{\hbar^{2}}{2m}\frac{\partial^{2}}{\partial x^{2}}+V_{\mathrm{R}}(x)+\hbar\omega_{\perp}+\\+2\hbar\omega_{\perp}a_{s}(N-1)|\psi_{0}(x)|^{2}\biggr]\psi_{0}(x)=\mu\psi_{0}(x) \,.
\end{multline}
The nonlinear coupling appearing in the last term of the Schrödinger operator is characterized hereafter by the constant 
\begin{equation}
\label{eqn:154}
g:=2\hbar\omega_{\perp}a_{s}\, .
\end{equation}
The random potential $V_R$ is a correlated speckle potential, typically engineered by exploiting the coupling between the atomic dipole moment and the electric field generated by shining coherent light on a diffusive plate~\cite{CleVarRet:6:2006}. Owing to the central limit theorem, both the real and the imaginary parts of the electric field in the observation point, for a high number of scattering events~\cite{RobertKuhn:Thesis:2007}, follow a Gaussian probability distribution. This leads to the formation of the speckle pattern where the probability distribution of the random potential amplitude is given by a Rayleigh law,
\begin{equation}
\label{eqn:055}
P(V_{\mathrm{R}})=\frac{\Theta_{H}(V_{\mathrm{R}}/V_{0})}{V_{0}}\,\mathrm{e}^{-V_{\mathrm{R}}/V_{0}} \,,
\end{equation}
$\Theta_{H}(x)$ being the Heaviside step function and $V_{0}$ the disorder strength. $V_0$ is inversely proportional to the detuning between the laser frequency and the atomic transition frequency~\cite{FalForIng:2008}. It is positive for blue-detuned speckles or negative for red-detuned ones. The spatial correlation profile $C(x)$ of the potential is chosen to be Gaussian, as one of those used in Ref.~\cite{Delande:MobEdgeSpeckle:PRL2014}: 
\begin{equation}
\label{eqn:056}
C(x):=\overline{\left[V_{\mathrm{R}}(0)-V_{0}\right]\left[V_{\mathrm{R}}(x)-V_{0}\right]}=V_{0}^{2}\mathrm{e}^{-\frac{x^{2}}{2\sigma^{2}}}\hspace{0.05cm}\text{,}
\end{equation}
in which $\sigma$ denotes the correlation length. The symbol $\bar{\cdot}$ indicates the ensemble average over all configurations of the disordered potential. 

For a 1D domain of length $L$ ($-L/2 \le x \le L/2$), the energy associated to the ground state of the GPE~\eqref{eqn:054} is given by~\cite{Stringari:BECRevTh:RMP99}
\begin{equation}
\begin{split}
\label{eqn:0071}
E_0 = & \int\limits_{-L/2}^{L/2} \biggl[\frac{\hbar^{2}}{2m}\biggl|\frac{\mathrm{d} \psi_0(x)}{\mathrm{d} x}\biggr|^{2}+\left( V_{\mathrm{R}}(x)+\hbar\omega_{\perp} \right) \bigl|\psi_{0}(x)\bigr|^{2}+\\&+\frac{g(N-1)}{2}\bigl|\psi_{0}(x)\bigr|^{4}\biggl]\hspace{0.025cm}\mathrm{d}x \,,
\end{split}
\end{equation}
whereas the corresponding chemical potential reads
\begin{equation}
\label{eqn:0072}
\mu=E_{0}+\frac{g(N-1)}{2}\int\limits_{-L/2}^{L/2}\bigl|\psi_{0}(x)\bigr|^{4}\hspace{0.025cm}\mathrm{d}x \,.
\end{equation}

\subsection{The localization landscape}
\label{subsec:locland}

In order to understand the behavior of the 1D ground state $\psi_{0}$ of the GPE, we start from the SP Hamiltonian $H^{\mathrm{sp}}$:
\begin{equation} 
\label{eqn:3157}
H^{\mathrm{sp}}:=-\frac{\hbar^{2}}{2m}\frac{\partial^{2}}{\partial x^{2}}+\hbar\omega_{\perp}+V_{\mathrm{R}}(x)\,.
\end{equation}
The localization landscape, introduced in Ref.~\cite{FilocheMayboroda:PNAS2012}, is then defined as the solution to
\begin{equation}
\label{eqn:902}
H^{\mathrm{sp}}u=1 \,.
\end{equation}
In this article, we choose to impose the Dirichlet boundary conditions on the LL (but they could be as well periodic, since they play no real role on localization effects):
\begin{equation} 
\label{eqn:903}
u(x)|_{x=\pm\frac{L}{2}} =0 \,.
\end{equation}
By decomposing an eigenstate $\psi^{\mathrm{sp}}$ of $H^{\mathrm{sp}}$ as $\psi^{\mathrm{sp}} = u\varphi^{\mathrm{sp}}$, where $\varphi^{\mathrm{sp}}$ is an auxiliary function and using~\eqref{eqn:902}, the time-independent Schrödinger equation $H^{\mathrm{sp}}\psi^{\mathrm{sp}}=E^{\mathrm{sp}}\psi^{\mathrm{sp}}$ can be rewritten as
\begin{equation} 
\label{eqn:3176}
-\frac{\hbar^{2}}{2m}\biggl[\frac{1}{u^{2}}\frac{\partial}{\partial x}\biggl(u^{2}\frac{\partial}{\partial x}\varphi^{\mathrm{sp}}\biggr) \biggr]+\frac{1}{u}\,\varphi^{\mathrm{sp}}=E^{\mathrm{sp}}\varphi^{\mathrm{sp}} \,.
\end{equation}

We see that the auxiliary function obeys a Schrödinger-like equation in which the Laplacian is replaced by a slightly more complicated elliptic operator whereas the original potential $V(x):=V_{\mathrm{R}}(x)+\hbar\omega_{\perp}$ is substituted by the effective potential $V_{\rm{LL}}(x)$, defined as
\begin{equation}
\label{eqn:884}
V_{\rm{LL}}(x):=u(x)^{-1} \,.
\end{equation}
It has been shown that this rewriting of the Schrödinger equation allows one to view the exponential localization (Anderson localization) of the lowest lying-states as a semiclassical confinement process in the smoother disordered potential $V_{\rm{LL}}$~\cite{FilocheMayboroda:PNAS2012, Arnold:EffConfLL:2016}. The LL also permits one to identify the position of those states without solving the full eigenvalue problem~\cite{Arnold:ComputingSpectraLL:2019} and accounts for the behavior of the tails of the integrated density of states~\cite{DesforgesFiloche:IDOS:2020}.

For the sake of simplicity, in the following treatment all quantities will be nondimensionalized, based on the units of the correlation function of the random potential. Hence, the lengths will be expressed in units of the correlation length $\sigma$ and the energies in units of the correlation energy $E_{\sigma}:=\frac{\hbar^{2}}{m\sigma^{2}}$~\cite{Delande:MobEdgeSpeckle:PRL2014}, which represents the zero-point energy for a particle confined in a spatial region of size equal to the correlation length $\sigma$.

\subsection{Numerical methods}
\label{subsec:nummeth}

For computing the ground state of the GPE, we adopt a Crank-Nicolson method introduced by Muruganandam and Adhikari~\cite{Adhikari:CodesGPE_CrankNicolson_SingleCPU:2009}. This method is based on the iteration an imaginary-time evolution process, performed by first considering only the potential and the nonlinear terms, then by involving the kinetic term of the GPE \eqref{eqn:054} which is discretized to the second order in the grid step~$\Delta x$.
The initial wavefunction is taken equal to the ground state $\psi_{0}^{\rm{sp}}$ of the noninteracting Hamiltonian $H^{\mathrm{sp}}$ and it is computed by solving the eigenvalue problem 
\begin{equation} 
\label{eqn:3057}
H^{\mathrm{sp}}\psi_{i}^{\mathrm{sp}}=E_{i}^{\mathrm{sp}}\psi_{i}^{\mathrm{sp}}\,\text{,}
\end{equation}
by using a divide-and-conquer algorithm for the diagonalization of symmetric matrices.
Starting from this SP state, the evaluation of $\psi_{0}$ proceeds as detailed in Ref.~\cite{Adhikari:CodesGPE_CrankNicolson_SingleCPU:2009}, with an intermediate solution at the subsequent time-step computed by retaining only the kinetic term. The potential as well as the nonlinear term are then introduced in a first-order time integration with the aim of achieving the complete evolution after a single time-step. The bipartite procedure thus outlined is repeated $N_{\mathrm{pas}}\sim 10^{5}$ times, a value which can be tuned to check the convergence of the result. The stationary state of the GPE thus obtained vanishes at the boundaries of the 1D domain.

Besides, in the computation of the energy associated to $\psi_{0}$ and of the corresponding chemical potential, both the integrations in Eqs~\eqref{eqn:0071} and~\eqref{eqn:0072} are performed numerically, using Cavalieri-Simpson's $3/8$ rule~\cite{Mathews:NumMeths:1991}.

The landscape function is calculated from Eq.~\eqref{eqn:902} by using a finite-element method and applying the boundary conditions shown in Eq.~\eqref{eqn:903}.

\section{\label{sec:attint}Attractive interactions}

The spatial behavior of the ground state of the GPE is first investigated in the case of attractive interactions. To this end, we examine the effect of both interactions and disorder on $\psi_{0}$. In the numerical simulations, we deal with samples of atoms with the same transverse-confinement length $l_{\perp}:=\sqrt{\frac{\hbar}{m\omega_{\perp}}}$ (equal to $11.1$ in our case) as the one of the harmonic potential used in the experiment of Ref.~\cite{LepFouBois:7:2016}. For the one-dimensional random potentials, we consider blue-detuned speckle potentials endowed with the same correlation length as in Ref.~\cite{Billy:AndersonBEC1D:N08}.

As it was first pointed out by Cheng and Adhikari in Ref.~\cite{ChengAdhikari:GPEstatsolnum:2010}, for attractive interactions the ground state of the GPE is localized in space and its tails decay exponentially. In finite quasi-1D systems, the left and the right tails do not exhibit exactly the same decay, which means that the modulus of the wavefunction can be approximated by:
\begin{equation}
\label{eqn:0084}
|\psi_{0}(x)| \approx c_{a} \, \begin{dcases}\mathrm{e}^{\frac{(x-x_{0})}{\lambda_{L}}}& -\frac{L}{2}\leq x <x_{0} \\
\mathrm{e}^{\frac{-(x-x_{0})}{\lambda_{R}}}& x_{0}\leq x\leq\frac{L}{2} \\
\end{dcases} \,,
\end{equation} 
where $x_{0}$ is the localization center, and $\lambda_{L}$ and $\lambda_{R}$ denote the left and the right localization length, respectively. Moreover, in Eq.~\eqref{eqn:0084}, $c_{a}$ represents the normalization coefficient.

\begin{figure}
\centering 
    \includegraphics[width=0.975\columnwidth]{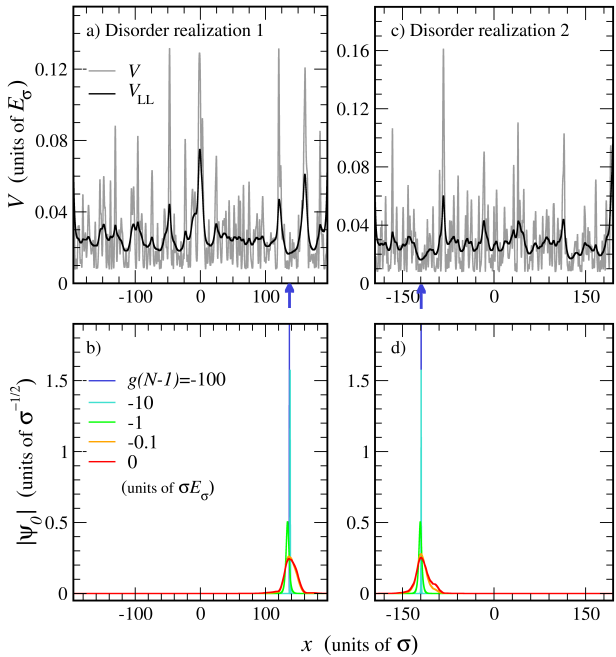} 
    \caption{Variable attractive interactions for Bose gases in two realizations of the blue-detuned speckle potential with $V_{0}=0.02$, as in Ref.~\cite{Billy:AndersonBEC1D:N08}, in a domain of length $L=40,000\Delta x$, where $\Delta x=0.01$. Panels (a) and (c): total potential $V$ (gray solid line) and effective potential $V_{\rm{LL}}$ (black solid line) along $Ox$. The blue arrows pinpoint the position of the absolute minima of the effective potentials. Panels (b) and (d): moduli of the ground states of the one-dimensional GPE (solid lines) for five different values of the coefficient of the nonlinear term and the disorder configurations displayed in Panels (a) and (c).}
\label{fig:EffPotWF_1D_Attr_Gvar_CfrDisReals}
\end{figure}

We first numerically determine the spatial behavior of the GS $\psi_0$ for different values of the coefficient of the nonlinear term $g(N-1)$, which vary from $-0.1$ to $-100$ by multiplicative steps of 10. In addition, we compute the ground state for vanishing interactions, using the parameters detailed at the beginning of this section.
As it can be inferred from Fig.~\ref{fig:EffPotWF_1D_Attr_Gvar_CfrDisReals}, where two realizations of a speckle potential with $V_{0}=0.02$ are displayed, the wavefunctions are localized close to the absolute minima $x_{\min{V_{\rm{LL}}}}$ of the effective potentials. The position of the localization center $x_{0}$ in Eq.~\eqref{eqn:0084} can thus be approximated as:
\begin{equation}
\label{eqn:1087}
x_{0} \approx x_{\min{V_{\rm{LL}}}}\hspace{0.05cm}\text{.}
\end{equation}
The accuracy of this approximation is assessed by computing the average of $\lvert x_{0} - x_{\min{V_{\mathrm{LL}}}}\rvert$ over 20 realizations of the random potential. As shown in Tab.~\ref{tab:Attr_VarG_CfrMaxPsi-MinW}, this deviation decreases for increasing nonlinear coefficient when $|g(N-1)| >1$ and does not exceed the correlation length of the speckle potential. 
Besides, the mean localization length $\bar{\lambda}:=\overline{(\lambda_{L}+\lambda_{R})}/2$ of the wavefunctions also diminishes as $|g(N-1)|$ is increased, as it can be also noticed from both Tab.~\ref{tab:Attr_VarG_CfrMaxPsi-MinW} and Fig.~\ref{fig:EffPotWF_1D_Attr_Gvar_CfrDisReals}.
The energy $E_{0}$ and the chemical potential $\mu$ associated to the GS of the GPE for attractive interactions are always lower than the energy of the GS of the corresponding noninteracting problem with the same potential.

\begin{figure}
\centering
    \includegraphics[width=0.975\columnwidth]{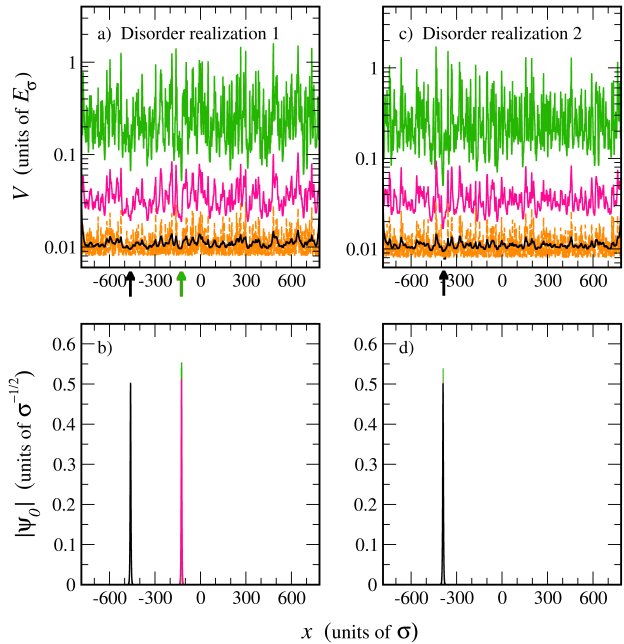}
   \caption{Variable disorder parameter for attractively interacting Bose gases in two realizations of the blue-detuned speckle potential. The domains are of length $L=40,000\Delta x$, where $\Delta x=0.04$. Panels (a) and (c): effective potential $V_{\rm{LL}}$ for $V_{0}=0.3$ (green solid lines), $V_{0}=0.03$ (magenta solid lines), $V_{0}=0.003$ (black solid lines). Total potential $V$ for $V_{0}=0.003$ (orange dashed line). The colored arrows pinpoint the positions of the absolute minima of the effective potentials above mentioned. Panels (b) and (d): ground state $\psi_{0}$ of the GPE with $g(N-1)=-1$ computed for the same values of $V_{0}$ as in Panels (a) and (c), represented by solid lines whose colors vary according to those of the effective potentials.}
\label{fig:EffPotWF_1D_Attr_V0var_CfrDisReals}
\end{figure}

\begin{table}[h!]
  \begin{center}
    \begin{tabular}{c|c|c|c|c}
      \toprule 
       $g(N-1)$ & $\overline{\lvert x_{0} - x_{\min{V_{\mathrm{LL}}}}\rvert}$ $\bigl[ \sigma \bigr]$ & $\overline{\lambda}$ $\left[\sigma\right]$ & $\overline{E_{0}}$ $\left[E_{\sigma}\right]$ & $\overline{\mu}$ $\left[E_{\sigma}\right]$\\
        &&&&\\ [-0.8em]
      \midrule
       0.0 & 0.388 & 8.65 & 0.0193 & 0.0193\\
       0.1 & 0.404 & 6.49 & $0.0173$ & $0.0150$\\  
       1 & 0.954 & 1.87 & $-0.0297$ & $-0.116$\\  
       10 & 0.222 & 0.157 & $-4.15$ & $-12.4$\\  
       100 & 0.033 & 0.020 & $-416$ & $-1219$\\  
      \bottomrule 
    \end{tabular} 
    \caption{Properties of the GS of the GPE obtained for the same values of $V_{0}$ and $g(N-1)$ used in Figs.~\ref{fig:EffPotWF_1D_Attr_Gvar_CfrDisReals}a-\ref{fig:EffPotWF_1D_Attr_Gvar_CfrDisReals}d. The modulus of the deviations between the exact positions of the localization centers and the absolute minima of the effective potentials are also reported, averaged over 56 disorder configurations. $\overline{E_{0}}$ and $\overline{\mu}$ denote the disorder-averaged total energy and chemical potential associated the wavefunctions respectively.}
    \label{tab:Attr_VarG_CfrMaxPsi-MinW}
  \end{center}
\end{table}
Fixing the value of the nonlinear coefficient $g(N-1)=-1$, we compute the GS of the GPE for two different realizations of the random potential and three different values of the disorder parameter $V_{0}=\{0.003, 0.03, 0.3\}$. All simulations, whose effective potentials are displayed in Figs.~\ref{fig:EffPotWF_1D_Attr_V0var_CfrDisReals}a and \ref{fig:EffPotWF_1D_Attr_V0var_CfrDisReals}c, confirm the conjecture in Eq.~\eqref{eqn:1087}. In particular, in Fig.~\ref{fig:EffPotWF_1D_Attr_V0var_CfrDisReals}a, the absolute minimum of $V_{\rm{LL}}$ shifts from $x=-462$ to $x=-125$, as signalled by the colored arrows, as the mean value $V_{0}$ of the speckle potential is increased from $0.003$ to $0.03$. In Fig.~\ref{fig:EffPotWF_1D_Attr_V0var_CfrDisReals}c, the absolute minima of the effective potentials lie at the same positions on the $x$ axis instead. The localization lengths of the wavefunctions, whose disorder-averaged values are reported in Tab.~\ref{tab:Attr_VarDis_CfrMaxPsi-MinW}, decrease as $V_{0}$ increases, in qualitative agreement with the SP case~\cite{LifGrePas:1988}.

We also remark that, since the wavefunction on the transverse directions is assumed to occupy the ground state of the 2D harmonic oscillator, the simulations on $\psi_{0}$ do not predict any collapse of the GS for strongly attractive interactions, in accordance with the case of square-well potentials~\cite{CarLeuRei:2000}.  Indeed, for higher-dimensional configurations, we expect the existence of a threshold in the nonlinear coefficient above which no stationary solution exists, as it was predicted in harmonic potentials~\cite{DodEdwWil:7:1996}. Compared to the study presented in Ref.~\cite{ChengAdhikari:GPEstatsolnum:2010}, the ground state of the GPE is here computed for a much wider range of nonlinear coefficients and disorder parameters. 
While the localization landscape allows to predict the position of the localization center of the wavefunctions, it is not able to account for the behavior of the localization length. An analytical description of the quantity last mentioned as a function of both the nonlinear coefficient and the disorder parameter would be of interest but lies beyond the scope of the paper.

\begin{table}[h!]
  \begin{center}
    \begin{tabular}{c|c|c|c|c}
      \toprule 
       $V_{0}$ & $\overline{\lvert x_{0} - x_{\min{V_{\mathrm{LL}}}}\rvert}$ $\bigl[ \sigma \bigr]$ & $\overline{\lambda}$ $\left[\sigma\right]$ & $\overline{E_{0}}$ $\left[E_{\sigma}\right]$ & $\overline{\mu}$ $\left[E_{\sigma}\right]$\\
        &&&&\\ [-0.8em]
      \midrule
       0.003 & 3.03 & 1.99 & $-0.0330$ & $-0.117$\\  
       0.03 & 0.979 & 1.82 & $-0.0299$ & $-0.116$\\  
       0.3 & 0.303 & 1.17 & $-0.0079$ & $-0.111$\\  
      \bottomrule 
    \end{tabular} 
    \caption{Properties of the GS of the GPE obtained for the same values of $V_{0}$ and $g(N-1)$ used in Figs.~\ref{fig:EffPotWF_1D_Attr_V0var_CfrDisReals}a-\ref{fig:EffPotWF_1D_Attr_V0var_CfrDisReals}d. The table shares the same structure as Tab.~\ref{tab:Attr_VarG_CfrMaxPsi-MinW}, and the disorder-averaged quantities are based on 56 configurations of the speckle potential.}
    \label{tab:Attr_VarDis_CfrMaxPsi-MinW}
  \end{center}
\end{table}

\section{\label{sec:repint}Repulsive interactions}

Let us now investigate the case of repulsive interactions, considering a Bose gas with a s-wave scattering length $a_{s}=0.02$~\cite{BaymPeth:1996} and transverse-confinement length $l_{\perp}=5.0$ in 1D correlated speckle potentials, as in the experiment of Billy \textit{et al.}~\cite{Billy:AndersonBEC1D:N08}.
In this section we first investigate the behavior of the ground state $\psi_{0}$ of the GPE as the nonlinear coefficient is varied, treating the regimes of weak interactions and intermediate or strong interactions separately. We later dwell on the shape and the properties of $\psi_{0}$ as the mean value $V_{0}$ of the random potential is varied, with a particular attention on the Lifshitz glass phase.

\subsection{Exploring the strength of the interactions}
\label{subsec:repvarint}

\begin{figure}
\centering
    \includegraphics[width=0.975\columnwidth]{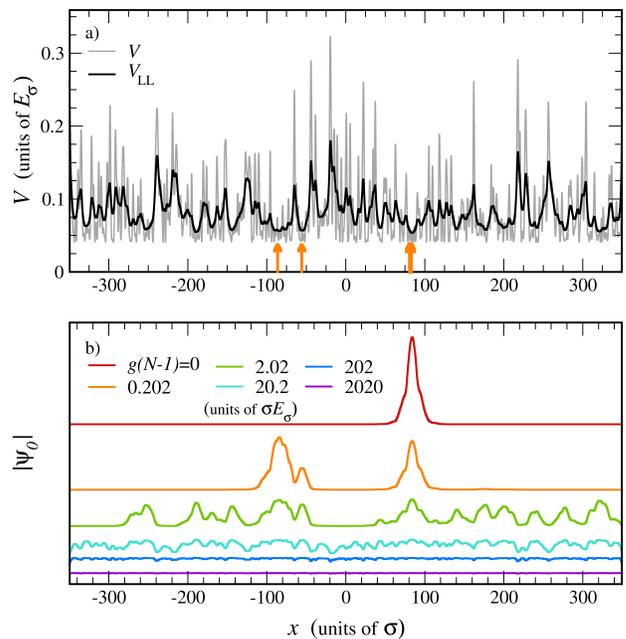}
   \caption{Variable repulsive interactions. Panel (a): speckle potential $V$ (grey solid line) with $V_{0}=0.044$ and effective potential $V_{\rm{LL}}$ (black solid line) in a domain of length $L=40,000\Delta x$, with grid step $\Delta x=0.0175$. The thick orange arrow indicates the absolute minimum of $V_{\rm{LL}}$, whereas the thin orange arrows pinpoint the three next-to-lowest minima of $V_{\rm{LL}}$. Panel (b): modulus of the ground state $|\psi_{0}|$ computed for different values of the nonlinear coefficient (solid lines): $g(N-1)=0$ (red), $=0.202$ (orange), $=2.02$ (green), $=20.2$ (turquoise), $=202$ (blue), $=2,020$ (violet). The baselines of the wavefunctions are shifted in order to display more clearly the curves.}
\label{fig:EffPotWF_1D_Num_Gvar}
\end{figure}

In order to provide an overview of the features of the ground state for repulsive interactions, we first plot it in Fig.~\ref{fig:EffPotWF_1D_Num_Gvar} for $g(N-1)=0.202\cdot10^{0},0.202\cdot10^{1},\dots,0.202\cdot10^{4}$, considering a single realization of the random potential, characterized by a mean value $V_{0}=0.044$, as in Ref.~\cite{Billy:AndersonBEC1D:N08}.
Starting from these data, we represent the original potential $V$ and the effective potential $V_{\rm{LL}}$ in Fig.~\ref{fig:EffPotWF_1D_Num_Gvar}a and the wavefunctions obtained in the noninteracting case as well as for the aforementioned five different values of $g(N-1)$ above mentioned in Fig.~\ref{fig:EffPotWF_1D_Num_Gvar}b.

As one increases the strength of the repulsive interaction, $\psi_{0}$ becomes significant in larger regions of the domain, eventually spreading over the whole interval $[-L/2,L/2]$. At the same time, the oscillation amplitude of the wavefunction decreases with increasing $g(N-1)$. 
 
In the noninteracting case, $\psi_{0}$ is exponentially localized, in accordance with theoretical predictions~\cite{LugAspSaP:7:2009} and experimental results~\cite{Billy:AndersonBEC1D:N08}. For the configuration displayed in Fig.~\ref{fig:EffPotWF_1D_Num_Gvar}a, the ground state (red curve in Fig.~\ref{fig:EffPotWF_1D_Num_Gvar}b) is localized at $x\approx 85$ which  corresponds to the absolute minimum of the effective potential $V_{\rm{LL}}$ (thick orange arrow in Fig.~\ref{fig:EffPotWF_1D_Num_Gvar}a), in agreement with the LL theory~\cite{Arnold:ComputingSpectraLL:2019}. $\psi_{0}$ decays exponentially in space and its localization length $\lambda:=(\lambda_{L}+\lambda_{R})/2$ amounts to $4.2$.

For $g(N-1)=0.202$, the ground state is significant only in two disconnected regions, as displayed in Fig.~\ref{fig:EffPotWF_1D_Num_Gvar}b. This indicates that the bosons fall into three clusters around the four lowest minima of the effective potential, pinpointed by the three orange arrows in Fig.~\ref{fig:EffPotWF_1D_Num_Gvar}a. For $g(N-1) \lesssim 2$, the Bose gas is fragmented into multiple regions, but does not explore the entire domain.

For $g(N-1) \gtrsim 20$, $\psi_{0}$ spreads over all the space (see Fig.~\ref{fig:EffPotWF_1D_Num_Gvar}b).
As pointed out in Tab.~\ref{tab:Rep_VarInt_CfrMaxMinPsiVW}, the healing length of the solutions to the GPE is smaller than the localization length $\lambda$ of the SP states endowed with energy equal to the chemical potential of $\psi_{0}$, except for the noninteracting case. For $g(N-1)~\gtrsim~10^{2}$, the localization length of the SP states is larger or comparable to $L$ and the states are thus delocalized in the entire domain in Fig.~\ref{tab:Rep_VarInt_CfrMaxMinPsiVW}.

\begin{table}[h!]
  \begin{center}
    \begin{tabular}{c|c|c|c|c|c|c}
      \toprule 
       $g(N-1)$  & $\xi$  & $\lambda$  & $D_{V}^{m}$  & $D_{V}^{M}$ 
       & $D_{V_{\rm{LL}}}^{m}$  & $D_{V_{\rm{LL}}}^{M}$ \\
        $\left[\sigma E_{\sigma}\right]$ & $\bigl[ \sigma \bigr]$ & $\left[\sigma\right]$  & $\left[\sigma\right]$ & $\left[\sigma\right]$ & $\left[\sigma\right]$ & $\left[\sigma\right]$\\ 
      \midrule
       0.0 & 14.02 & 4.2 & - & - & - & -\\  
       0.202 & 2.75 & 7.7 & - & - & - & -\\  
       2.02 & 2.6 & 7.3 & - & - & - & -\\  
       20.2 & 2.14 & 24 & 3.1 & 2.9 & 0.65 & 0.79\\ 
       202 & 1.16 & - & 1.3 & 1.2 & 0.25 & 0.5\\
       2,020 & 0.41 & - & 0.45 & 0.60 & 0.33 & 0.90\\
      \bottomrule 
    \end{tabular} 
    \caption{Properties of the GSs displayed in Fig.~\ref{fig:EffPotWF_1D_Num_Gvar}. In the first two columns, the healing length $\xi$ and the localization length $\lambda$ of the SP state with energy equal to the chemical potential of the GS of the GPE are reported. In the two following columns, the average distance $D_{V}^{m}$ (resp. $D_{V}^{M}$) between the local minima (resp. maxima) of $\psi_{0}$ and the closest local maxima (resp. minima) of $V$ is reported. The last two columns contain the last mentioned quantities computed for the effective potential $V_{\rm{LL}}$, and denoted $D_{V_{\rm{LL}}}^{m}$ and $D_{V_{\rm{LL}}}^{M}$. The average distances are not computed for the GSs with $g(N-1) <10$ since these states are not delocalized in the whole domain.}
    \label{tab:Rep_VarInt_CfrMaxMinPsiVW}
  \end{center}
\end{table}

In addition, for $20 \lesssim g(N-1) \lesssim 10^{3} $ the maxima and minima of $\psi_{0}$ occur closer to the minima and maxima respectively of the effective potential $V_{\rm{LL}}$ than to those of the original potential $V$.  This is illustrated in Tab.~\ref{tab:Rep_VarInt_CfrMaxMinPsiVW}, where the average distances $D_{V_{\rm{LL}}}^{m}$ (resp. $D_{V_{\rm{LL}}}^{M}$) between the minima (resp. maxima) of the wavefunction and the nearest maxima (resp. minima) of the $V_{\rm{LL}}$ are compared to those computed starting from $V$, denoted as $D_{V}^{m}$ (resp. $D_{V}^{M}$).

For $g(N-1)\gtrsim 10^{3}$, the wavefunction follows more accurately the modulations of the original potential $V(x)$ instead and is predicted by the Thomas-Fermi approximation~\cite{SanchezPalencia:Disorder1DGPEThomasFermi:2006}, as long as the healing length is smaller than the correlation length $\sigma$.
The overlap integral between the numerical wavefunction and the one in Thomas-Fermi approximation for the last-mentioned value of $g(N-1)$ amounts in fact to $1.000$. 

In the limit of infinite repulsive interactions the system is ergodic~\cite{FalNatPok:2009} and $|\psi_{0}|$ is nearly constant in space, and thus independent of the disorder realization. It is worth noting that, in this limit, the Gross-Pitaevskii mean-field approch is not valid anymore, since the constraint on the mean particle density $\langle \rho \rangle |a_{s}|^{3} \ll 1$ is not satisfied. The GS of gas of hardcore bosons is then described by the Tonks-Girardeau model, according to which the gas behaves as a system of spinless fermions in the random potential $V$~\cite{SeiWar:2016}.

\subsubsection{Weak interactions}
\label{subsubsec:wearepint}
As it was pointed out in Ref.~\cite{Kivshar:PertGPE:2001} where a pure harmonic potential was studied, for weak repulsive interactions, the spatial behavior of the ground state of the GPE can be understood through the lowest-lying eigenstates of the noninteracting problem in Eq.~\eqref{eqn:3057}. We will later show that these eigenstates can be approximated using the LL theory.

For a low nonlinear coefficient, $0 < g(N-1) \lesssim 1$, the ground state $\psi_{0}$ of the GPE can be indeed expressed as a linear combination of the $N_{s}$ eigenstates of $H^{\mathrm{sp}}$ whose energy does not exceed $E^{\rm{th}}$. This energy threshold is defined as: 
\begin{equation}
\label{eqn:0091}
E^{\rm{th}}:=E_{0}^{\mathrm{sp}}+\frac{g(N-1)}{2}\int\limits_{-L/2}^{L/2}|\psi_{0}^{\mathrm{sp}}(x)|^{4}\hspace{0.05cm}\mathrm{d}x\hspace{0.05cm}\text{.}
\end{equation}
The number of SP states $N_{s}$ contributing to the expression for $\psi_{0}$ thus satisfy
\begin{equation} 
\label{eqn:9086}
N_{s} = n^{\mathrm{sp}}\bigl(E^{\rm{th}}\bigr)\hspace{0.05cm}\text{,}
\end{equation}
where $n^{\mathrm{sp}}$ indicates the integrated density of SP states (IDoS) evaluated at $E^{\rm{th}}$. 

The ground state of the GPE can be then written as: 
 \begin{equation}
 \label{eqn:0086}
 \psi_{0}(x) \approx \sum\limits_{j=0}^{N_{s}-1}c_{j}\psi_{j}^{\mathrm{sp}}(x)\hspace{0.05cm}\text{,}
 \end{equation}
where the coefficients $\{c_{j}\}$ must satisfy $\sum_{j=0}^{N_{s}-1}|c_{j}|^{2}=1$. While in the harmonic case the coefficients related to odd eigenfunctions vanish due to the parity symmetry of the potential, here, since the speckle potential lacks any spatial symmetry, the $\{c_{j}\}$ of the lowest-energy states can be all nonzero and are evaluated as detailed in App.~\hyperref[app:coeffeval]{A}.

Labeling as $\psi_{0}^{\mathrm{lcs}}$ the state approximated using the \emph{linear combination of SP states} in Eq.~\eqref{eqn:0086}, the total energy of the gas $E_{0}^{\mathrm{lcs}}$ can be evaluated in this framework by inserting the right-hand side of Eq.~\eqref{eqn:0086} into Eq.~\eqref{eqn:0071}, thus obtaining:
\begin{equation} 
\begin{split}
\label{eqn:285}
E_{0}^{\mathrm{lcs}}:=&\sum_{j=0}^{N_{s}-1}|c_{j}|^{2}E_{j}^{\mathrm{sp}}+\\&+\frac{g(N-1)}{2}\int\limits_{-L/2}^{L/2}\Biggl|\sum_{j=0}^{N_{s}-1}c_{j}\psi_{j}^{\mathrm{sp}}(x)\Biggr|^{4}\hspace{0.05cm}\mathrm{d}x\hspace{0.05cm}\text{.}
\end{split}
\end{equation}
By labeling the coefficients of the decomposition in SP states as $\{c_{i}^{\rm{LL}}\}$ and by plugging the right-hand side of Eq.~\eqref{eqn:0086} into Eq.~\eqref{eqn:0072}, the chemical potential can be analogously expressed as:
\begin{equation}
\begin{split}
\label{eqn:0286}
\mu^{\rm{lcs}}:=&\sum_{j=0}^{N_{s}-1}|c_{j}|^{2}E_{j}^{\mathrm{sp}}+\\&+g(N-1)\int\limits_{-L/2}^{L/2}\Biggl|\sum_{j=0}^{N_{s}-1}c_{j}\psi_{j}^{\mathrm{sp}}(x)\Biggr|^{4}\hspace{0.05cm}\mathrm{d}x\hspace{0.05cm}\text{.}
\end{split}
\end{equation}
The SP eigenstates, necessary for computing $\psi_{0}^{\rm{lcs}}$, $E_{0}^{\mathrm{lcs}}$ and $\mu^{\rm{lcs}}$ can also be efficiently computed by starting from the LL.

The correspondence between the position of the lowest minima of the effective potential and the localization centers of the lowest-lying SP states has been indeed illustrated in Ref.~\cite{Arnold:ComputingSpectraLL:2019,ArnDavFil:5:2019}. In the former work, it is also proved that the domain $\Omega_{i}$ of the $i$-th lowest-lying SP state can be identified as the connected region where $V_{\rm{LL}}(x)\leq E_{i}^{\mathrm{sp,LL}}$, containing the $i$-th lowest minimum of $V_{\rm{LL}}$.

The energy $E_{i}^{\mathrm{sp,LL}}$ of the lowest-lying SP states can be estimated using an empirical formula introduced in Ref.~\cite{Arnold:ComputingSpectraLL:2019}: 
\begin{equation}
\label{eqn:0085}
E_{i}^{\mathrm{sp,LL}}=\biggl(1+\frac{d}{4}\biggr)V_{\mathrm{LL}\hspace{0.025cm}\mathrm{min},i}\hspace{0.05cm}\text{,}
\end{equation}
where $V_{\mathrm{LL}\hspace{0.025cm}\mathrm{min},i}$ now denotes the \textsl{absolute} minimum of the effective potential in the domain $\Omega_{i}$. The SP wavefunction of the $i$-th excited state, whose support lies in $\Omega_{i}$, can be expressed as
\begin{equation}
\label{eqn:0075}
\psi_{i}^{\mathrm{sp,LL}}(x)=\frac{u(x)}{\bigl(\int_{\Omega_{i}}|u(x)|^{2}\hspace{0.05cm}\mathrm{d}x\bigr)^{1/2}}\hspace{0.05cm}\textit{,}
\end{equation}
where $u$ is the localization landscape, defined in Eq.~\eqref{eqn:902}. The soundness of Eq.~\eqref{eqn:0075}, which was introduced in Ref.~\cite{Arnold:ComputingSpectraLL:2019}, is also discussed in App.~\hyperref[app:loclandSPstates]{B}.

Eqs.~\eqref{eqn:0085}-\eqref{eqn:0075} are valid as long as the wells of the effective potential are occupied by a single SP state. This situation is by far the most common in blue-detuned speckle potentials, owing to the particular form of the probability distribution in Eq.~\eqref{eqn:055}, which takes its maximum value for $V_{\rm{R}}\equiv 0$, at the lower bound of its domain.

By plugging Eqs.~\eqref{eqn:0085} and~\eqref{eqn:0075} into~\eqref{eqn:0091}, one can also find the energy threshold for the SP states within the LL approach:
\begin{equation}
\label{eqn:0087}
E^{\rm{th,LL}}:=\biggl(1+\frac{d}{4}\biggr)V_{\mathrm{LL}\hspace{0.025cm}\mathrm{min},0}+\frac{g(N-1)}{2}\frac{\mathcal{J}_{4}^{\rm{sp}}}{(\mathcal{J}_{2}^{\rm{sp}})^{2}}\hspace{0.05cm}\text{,}
\end{equation}
where $\mathcal{J}_{l}^{\rm{sp}}$ labels the integral:
\begin{equation}
\label{eqn:0088}
\mathcal{J}_{l}^{\rm{sp}} := \int \limits_{\Omega_{0}}|u(x)|^{l}\hspace{0.05cm}\mathrm{d}x\hspace{0.05cm}\textit{,}\quad\quad\quad l=2,4\,.
\end{equation}

Coherently with the analysis carried out so far, the same quantities are also evaluated by using the LL.
By taking advantage of Eq. \eqref{eqn:0085}, the total energy can be in fact expressed as
 \begin{equation}
\begin{split}
\label{eqn:085}
&E_{0}^{\mathrm{lcs,LL}}:=\sum_{j=0}^{N_{s}^{\rm{LL}}-1}\biggl(1+\frac{d}{4}\biggr)V_{\rm{LL}\hspace{0.025cm}\rm{min},j}|c^{\mathrm{LL}}_{j}|^{2}+\\&+
\frac{g(N-1)}{2(\mathcal{J}_{2}^{\rm{sp}})^{2}}\sum_{i,j,k,l=0}^{N_{s}^{\rm{LL}}-1}c_{i}^{\rm{LL}\hspace{0.025cm}*}c_{j}^{\rm{LL}\hspace{0.025cm}*}c_{k}^{\rm{LL}}c_{l}^{\rm{LL}}\int_{\vspace{-0.1cm}\Omega_{i}\cap\Omega_{j}\cap\Omega_{k}\cap\Omega_{l}}\hspace{-1.6cm}|u(x)|^{4}\mathrm{d}x\hspace{0.05cm}\text{,}
\end{split}
\end{equation}
where $N_{s}^{\rm{LL}}$ is the number of SP states whose energy is lower than $E^{\rm{th,LL}}$, defined in Eq.~\eqref{eqn:0087}.
Similarly to Eq.~\eqref{eqn:0286}, the chemical potential in the LL approximation can be written as:
\begin{equation}
\begin{split}
\label{eqn:0287}
&\mu^{\mathrm{lcs,LL}}:=\sum_{j=0}^{N_{s}^{\rm{LL}}-1}\biggl(1+\frac{d}{4}\biggr)V_{\rm{LL}\hspace{0.025cm}\rm{min},j}|c^{\mathrm{LL}}_{j}|^{2}+\\&
+\frac{g(N-1)}{(\mathcal{J}_{2}^{\rm{sp}})^{2}}\sum_{i,j,k,l=0}^{N_{s}^{\rm{LL}}-1}c_{i}^{\rm{LL}\hspace{0.025cm}*}c_{j}^{\rm{LL}\hspace{0.025cm}*}c_{k}^{\rm{LL}}c_{l}^{\rm{LL}}\int_{\vspace{-0.1cm}\Omega_{i}\cap\Omega_{j}\cap\Omega_{k}\cap\Omega_{l}}\hspace{-1.6cm}|u(x)|^{4}\mathrm{d}x\hspace{0.05cm}\text{.}
\end{split}
\end{equation}

\begin{figure}
\centering 
\includegraphics[width=0.975\columnwidth]{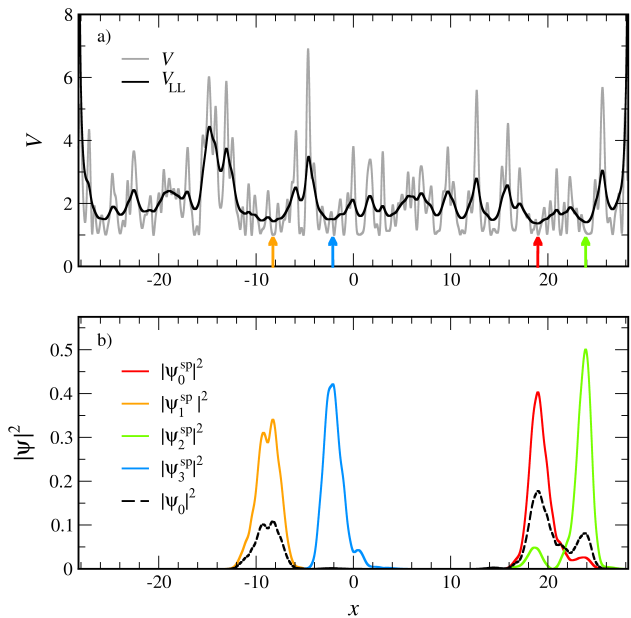} 
\caption{Weakly repulsive interactions and SP states. Panel (a): original potential $V$ and effective potential $V_{\rm{LL}}$ computed using the same values of $V_{0}$, $l_{\perp}$ and grid step $\Delta x$ as in Fig.~\ref{fig:EffPotWF_1D_Num_Gvar}, for a domain of length $L=16,000\Delta x$. The arrows indicate the four lowest minima of $V_{\rm{LL}}$ where the lowest-lying SP states reach their absolute maxima. Panel (b): probability amplitudes associated to the four lowest-energy states $\{\psi_{i}^{\mathrm{sp}}\}$ of the noninteracting Hamiltonian $H^{\rm{sp}}$, plotted as solid lines. The aforementioned states are found by exact diagonalization of the SP Hamiltonian in Eq.~\eqref{eqn:3157}. The black dashed line represents the ground state $\psi_{0}$ of the GPE, extracted for $g(N-1)=0.202$.}
\label{fig:CfrDiagImagTime_87Rb_G01}
\end{figure}

In order to numerically test the validity of this approach, we consider an ultracold Bose gas in a speckle potential with the same values of parameters $V_{0}$ and $\sigma$ as in Fig.~\ref{fig:EffPotWF_1D_Num_Gvar}, but a different disorder realization in a smaller domain, represented in Fig.~\ref{fig:CfrDiagImagTime_87Rb_G01}a. The ground state of the GPE is computed for $g(N-1)=0.202$ and plotted in Fig.~\ref{fig:CfrDiagImagTime_87Rb_G01}b by the black dashed lines respectively. For the former value of the nonlinear coefficient, the number of relevant SP eigenstates of $H^{\rm{sp}}$ contributing to the expansion~\eqref{eqn:0086} amounts to $N_{s}=4$, value that is reckoned by using Eqs.~\eqref{eqn:9086} and~\eqref{eqn:0091}. The four lowest-lying SP eigenstates, computed by exact diagonalization of the SP Hamiltonian, are plotted in Fig.~\ref{fig:CfrDiagImagTime_87Rb_G01}b (solid lines).

One can also infer from Fig.~\ref{fig:CfrDiagImagTime_87Rb_G01} that the absolute maxima of the SP eigenstates occur at the lowest minima of the effective potential $V_{\rm{LL}}$. In particular, the squared modulus of $\psi_{0}^{\rm{sp}}$ reaches its maximum at the absolute minimum of the effective potential $V_{\rm{LL}}$, as pinpointed by the red arrow in Fig.~\ref{fig:CfrDiagImagTime_87Rb_G01}a. The other SP states possess their absolute maxima at the local minima of $V_{\rm{LL}}$ indicated by the other arrows in the same figure.  
While the eigenstate $\psi_{2}^{\rm{sp}}$ is localized, with a localization length of $\lambda=6.3$, the ground state $\psi_{0}$ of the GPE, whose chemical potential is close to $E_{2}^{\rm{sp}}$, is partially delocalized and possesses a healing length $\xi=2.70$.

For the situation in Fig.~\ref{fig:CfrDiagImagTime_87Rb_G01}, we compute the coefficients $\{ c_{i}\}$ following a procedure detailed in App.~\hyperref[app:coeffeval]{A}, that is by solving Eq.~\eqref{eqn:0186} with $m=0,1,\dots,N_{s}-1$, which involves the SP states and energies computed by two different methods: exact diagonalization of $H^{\rm{sp}}$ and LL (in Eqs.~\eqref{eqn:0085} and~\eqref{eqn:0075}).
The moduli of the coefficients $\{c_{i}\}$ found with the former method are presented in Tab.~\ref{tab:Cfr_ExpCoeff_Estimates_Ng1}, together with the ones computed using the latter method, $\{ |c_{i}^{\rm{LL}}| \}$.
While the number of relevant SP states is not the same for the two approaches, the agreement between the two sets of coefficients is quite satisfactory, in particular for the main two contributions, $|c_{0}|$ and $|c_{1}|$, where the deviation is about $6\%$ on average. The discrepancy occurring for the most excited states follows from the coarseness of the LL-based approximation in Eq.~\eqref{eqn:0075} in the tail regions of the SP wavefunctions.
As one can see in Tab.~\ref{tab:Cfr_ExpCoeff_Estimates_Ng1}, the values of the energies computed by Eq.~\eqref{eqn:0085} are overestimated by $9\%$ on average, an amount which is of the same order of the one found in Ref.~\cite{Arnold:ComputingSpectraLL:2019}.

\begin{table}[h!]
  \begin{center}
    \begin{tabular}{l|c|c|c|c}
      \toprule 
       $i$ & $E_{i}^{\rm{sp}}$ $\left[10^{-2}E_{\sigma}\right]$ & $E_{i}^{\rm{sp,LL}}$ $\left[10^{-2}E_{\sigma}\right]$ & $|c_{i}|$ & $|c_{i}^{\rm{LL}}|$\\
        &&&\\ [-0.8em]
      \midrule
       0 & 6.412 & 6.937 & 0.766 & 0.674\\ 
       1 & 6.555 & 7.279 & 0.587 & 0.578\\ 
       2 & 6.848 & 7.124 & 0.248 & 0.460\\ 
       3 & 6.917 & 7.600 &  0.0823 & 0.0\\ 
      \bottomrule 
    \end{tabular} 
    \caption{Summary of the values of the energy and of the coefficients of the expansion~\eqref{eqn:0086} related to the state $\psi_{0}$ plotted in Fig.~\ref{fig:CfrDiagImagTime_87Rb_G01}b. $\{E_{i}^{\rm{sp}}\}$ and $\{|c_{i}|\}$ are computed by means of the eigenfunctions extracted by exact diagonalization of $H^{\rm{sp}}$. $\{E_{i}^{\rm{sp,LL}}\}$ and $\{|c_{i}^{\rm{LL}}|\}$ are evaluated by using the SP eigenstates in the LL-based approximation in Eq.~\eqref{eqn:0075}, with the eigen-energies in Eq.~\eqref{eqn:0085}.}
    \label{tab:Cfr_ExpCoeff_Estimates_Ng1}
  \end{center}
\end{table}

\begin{figure}
\centering
    \includegraphics[width=0.975\columnwidth]{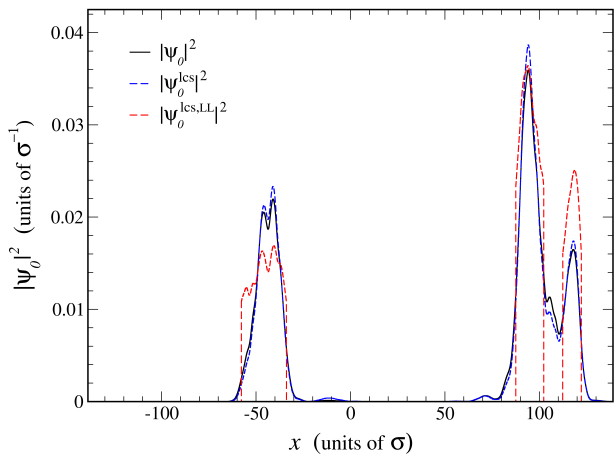}
   \caption{Probability density associated to the ground state of the GPE for $g(N-1)=0.202$ and the disorder configuration displayed in Fig.~\ref{fig:CfrDiagImagTime_87Rb_G01}a, computed by means of three different methods. Probability amplitude $|\psi_{0}|^{2}$ determined by imaginary-time evolution (black solid line). $|\psi_{0}^{\rm{lcs}}|^{2}$ (blue dashed line), referring to the approximation in Eq.~\eqref{eqn:0086}, based on the SP eigenstates computed by exact diagonalization. 
$|\psi_{0}^{\rm{lcs,LL}}|^{2}$ (red dashed line), found in the approximation in Eq.~\eqref{eqn:0086}, based on the SP eigenfunctions estimated using the LL in Eq.~\eqref{eqn:0075}.
}
\label{fig:WaveFx_Ng1_GPE-lcsp_CfrMeth}
\end{figure}

By using the two sets of coefficients in Tab.~\ref{tab:Cfr_ExpCoeff_Estimates_Ng1}, the ground state of the GPE in the approximation presented in Eq.~\eqref{eqn:0086} is evaluated using the SP eigenstates extracted by means of the two different approaches.
The probability amplitude $|\psi_{0}^{\mathrm{lcs}}|^{2}$ computed with the coefficients $\{ c_{i}\}$ is then represented in Fig.~\ref{fig:WaveFx_Ng1_GPE-lcsp_CfrMeth} as the blue dashed line, whereas the one obtained with the coefficients $\{c_{i}^{\rm{LL}}\}$, indicated as $|\psi_{0}^{\mathrm{lcs, LL}}|^{2}$, is plotted as the red dashed line. 
In the same figure, both quantities are compared against the exact numerical probability amplitude $|\psi_{0}|^{2}$ (black solid line).
The very good agreement between the squared moduli of $\psi_{0}$ and $\psi_{0}^{\rm{lcsp}}$ also ascertains the validity of the approach here used, based on the lowest-lying SP eigenstates. Moreover, the approximation in Eq.~\eqref{eqn:0086} is also accurate in the description of $\psi_{0}$ in the tail regions. In these spatial intervals, the nonlinear term of the GPE becomes negligible compared to the one-particle terms, so that the solutions to the GPE decay exponentially as the SP states.

The overlap integral between the wavefunctions $\psi_{0}$ and $\psi_{0}^{\rm{lcsp}}$, $\int \psi_{0}^{*}(x)\psi_{0}^{\mathrm{lcs}}(x)\hspace{0.05cm}\mathrm{d}x$, is equal to 0.996, while the overlap integral between $\psi_{0}$ and $\psi_{0}^{\mathrm{lcs,LL}}$ amounts then to 0.901.
This smaller value reflects a larger deviation between the states. 
On the other hand, $\psi_{0}^{\mathrm{lcs,LL}}$ is able to capture well the positions of the highest local maxima, but fails to account for the region where $\psi_{0}^{\rm{sp}}$ and $\psi_{2}^{\rm{sp}}$ overlap between each other. 
Nevertheless, the landscape-based approximation becomes convenient for larger systems or for higher dimension, due to its much lower computational cost~\cite{Arnold:ComputingSpectraLL:2019} compared to the diagonalization of the SP Hamiltonian.

Furthermore, the total energy and the chemical potential associated to the ground state of the GPE in the SP-state expansion of Eq.~\eqref{eqn:0086} are computed by means of Eqs.~\eqref{eqn:285} and~\eqref{eqn:0286}, respectively.
The values of the energy and the chemical potential thus found are compared with those obtained using the LL in Eqs.~\eqref{eqn:085} and ~\eqref{eqn:0287} in the second and the third columns of Tab.~\ref{tab:Cfr_Energy_Estimates_Ng1}, respectively. 

In particular, the energy and the chemical potential obtained by using the exact $\psi_{0}$ (see Eqs.~\eqref{eqn:0071} and~\eqref{eqn:0072}) appear to be in excellent agreement with those found within the approximation based on the expansion in SP states extracted by exact diagonalization of $H^{\rm{sp}}$, the deviation between the two estimates being of $3\permil$ at most. The discrepancies with the estimates based on the LL, $E_{0}^{\mathrm{lcs,LL}}$ and $\mu^{\mathrm{lcs,LL}}$ both amount to $10\%$ instead and mirrors the overestimates noticed in the SP energies (in Tab.~\ref{tab:Cfr_ExpCoeff_Estimates_Ng1}).
\begin{table}[h!]
  \begin{center}
    \begin{tabular}{c|c|c}
      \toprule 
       Exact solution & \multicolumn{2}{c}{SP-mode decomposition} \\ \hline
       &&\\ [-0.8em]
      $E_{0}$ $\left[E_{\sigma}\right]$ & $E_{0}^{\mathrm{lcs}}$ $\left[E_{\sigma}\right]$ & $E_{0}^{\mathrm{lcs,LL}}$  $\left[E_{\sigma}\right]$\\ 
       &&\\ [-0.85em]
      \midrule 
      $6.685\cdot 10^{-2}$ & $6.689\cdot 10^{-2}$ & $7.328\cdot 10^{-2}$\\ 
      \toprule
       $\mu$ $\left[E_{\sigma}\right]$ & $\mu^{\mathrm{lcs}}$ $\left[E_{\sigma}\right]$ & $\mu^{\mathrm{lcs,LL}}$ $\left[E_{\sigma}\right]$\\ 
       &&\\ [-0.85em]
      \midrule 
      $6.872\cdot 10^{-2}$ & $6.888\cdot 10^{-2}$ & $7.564\cdot 10^{-2}$\\  
      \bottomrule 
    \end{tabular}
    \caption{Comparison between the values of the energy and the chemical potential of the ground state of the GPE for $g(N-1)=0.202$ in Fig.~\ref{fig:WaveFx_Ng1_GPE-lcsp_CfrMeth}, obtained by three different procedures. First column: quantities computed by means of the Crank-Nicolson imaginary-time evolution algorithm. Second column: energy and chemical potential evaluated by using Eqs.~\eqref{eqn:285} and~\eqref{eqn:0286} respectively. Third column: same quantities estimated by making use of Eqs.~\eqref{eqn:085} and~\eqref{eqn:0287} respectively.}
    \label{tab:Cfr_Energy_Estimates_Ng1}
  \end{center}
\end{table}

Finally, it is worth noticing that the GSs of the GPE explainable as superpositions of SP states are not necessarily in the Lifshitz glass regime~\cite{LugCleBou:6:2007}, where the spatial overlap between the states is negliglible, unlike the case in Fig.~\ref{fig:CfrDiagImagTime_87Rb_G01}.

We have seeen that, for weak repulsive interactions, the delocalization effect can be understood by introducing a decomposition in the lowest-lying SP states (see Eq.~\eqref{eqn:0086}). This applies to any type of random potential, spatial distribution, and correlation profile. Besides, the LL, which is able to predict the location of each SP state, allows one to quite precisely estimate the ground state of the GPE, as well as its energy and chemical potential, with a reduced computational cost. A more accurate description of the long-distance behaviour of the SP states, and hence of the GS of the GPE, would be possible by taking advantage of the notion of Agmon's distance~\cite{Agmon:LN:1985,Arnold:EffConfLL:2016,ShaDylTho:4:2021}, which is also based on the effective potential $V_{\rm{LL}}$. This approach would be particularly relevant to assess the transport properties of $\psi_{0}$, which however lie outside the scope of the paper.
The evaluation of the SP ground state by exact diagonalization of $H^{\rm{sp}}$ has further allowed us to prove the consistency with the stationary state computed by the imaginary-time evolution algorithm for vanishing interactions. 
In the following subsection we focus on the shape of the wavefunction $\psi_{0}$ for intermediate and strong repulsive interactions.

\subsubsection{Intermediate and strong interactions}
\label{subsubsec:strrepint}

\begin{figure*}
\centering
    \includegraphics[width=0.975\linewidth]{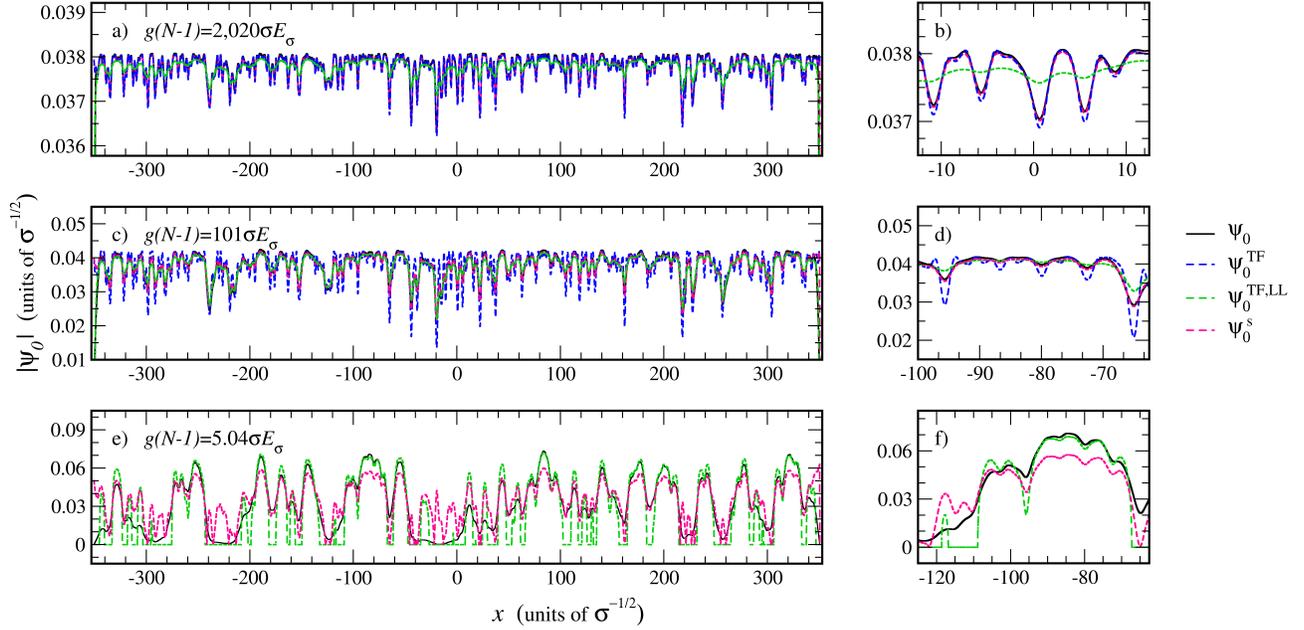}
\caption{Intermediate and strongly repulsive interactions. Panel (a): ground state of the GPE with the potential shown in Fig.~\ref{fig:EffPotWF_1D_Num_Gvar}a and for a nonlinear coefficient $g(N-1)=2,020$. Wavefunction $\psi_{0}$ computed by imaginary-time evolution (black solid line), by the Thomas-Fermi approximation $\psi_{0}^{\mathrm{TF}}$ (blue dashed line)  in Eq.~\eqref{eqn:2084}, by the lansdcape-based approximation $\psi_{0}^{\mathrm{TF,LL}}$ (green dashed line) and by the perturbative method in Eq.~\eqref{eqn:0070} $\psi_{0}^{s}$ (magenta dashed line). Panel (b): the same quantities as in Panel (a), restricted to the interval $[-12.5,12.5]$. Panel (c): the same quantities as in Panel (a), but related to the ground state of the GPE for $g(N-1)=101$. Panel (d): the same quantities as in Panel (c), restricted to the interval $[-100,-62.5]$. Panel (e): the same quantities as in Panel (a), but related to the ground state of the GPE for $g(N-1)=5.04$. Unlike in Panel (a), $\psi_{0}^{\mathrm{TF}}$ is here omitted, since the conditions of the Thomas-Fermi approximation are by far not satisfied. Panel (f): the same quantities as in Panel (e), restricted to the interval $[-125,-62.5]$.
}
\label{fig:WaveFx_HighNg_Rep_Approxes}
\end{figure*}

As proved in Ref.~\cite{SanchezPalencia:Disorder1DGPEThomasFermi:2006}, when $\xi \gtrsim 1$, the length of the spatial modulations of the particle density can be only larger than the correlation length of the random potential~$V_{\rm{R}}$. The wavefunction of the ground state $\psi_{0}$ is then sensitive to the modulations of a potential $V_{\rm{s}}$ which is smoother~\cite{SanchezPalencia:Disorder1DGPEThomasFermi:2006} than the original one, $V$:
\begin{equation}
\label{eqn:0069}
V_{\rm{s}}(x)=\int\limits_{-L/2}^{L/2}G(x')V_{\rm{R}}(x-x')\hspace{0.05cm}\mathrm{d}x'\hspace{0.05cm}\text{,}
\end{equation}
where 
\begin{equation}
\label{eqn:0099}
G(x)=\frac{1}{\sqrt{2}\xi'}\mathrm{e}^{-\frac{\sqrt{2}|x|}{\xi'}} 
\end{equation}
is the Green's function related to the disorder-free problem $(-\frac{\xi'^{2}}{2}\nabla^{2}+l_{\perp}^{-2})G(x)=\delta(x)$, in which $\xi':=\xi\sqrt{\frac{\mu}{\mu-l_{\perp}^{-2}}}$. 
The macroscopical wavefunction in this approximation, where the smoothed potential is treated as a perturbation with respect to the homogeneous case, $\psi_{0}^{s}$ is given by~\cite{SanchezPalencia:Disorder1DGPEThomasFermi:2006}:
\begin{equation}
\label{eqn:0070}
\psi_{0}^{s}(x)=\sqrt{\frac{\mu-l_{\perp}^{-2}}{g(N-1)}}\Biggl(1-\frac{1}{2(\mu-l_{\perp}^{-2})}V_{\rm{s}}(x)\Biggr)\hspace{0.05cm}\text{.}
\end{equation}
In addition, the validity of Eq.~\eqref{eqn:0070} is guaranteed as long as $\xi \ll L$ and $V_{\rm{s}}(x) \ll \mu -l_{\perp}^{-2}$. 

Within the smoothing regime thus defined, we introduce another approximation scheme, based on the effective potential $V_{\rm{LL}}(x)$ and much less computationally expensive:
\begin{equation}
\label{eqn:0083}
|\psi_{0}^{\mathrm{TF,LL}}(x)| = \begin{dcases}\sqrt{\frac{\mu-V_{\rm{LL}}(x)}{g(N-1)}}& \mu\geq V_{\rm{LL}}(x)\\
0 & \mu < V_{\rm{LL}}(x)\\
\end{dcases} \,.
\end{equation} 
This scheme is accurate as long as $1 < \xi < \sigma_{\rm{LL}}$, where $\sigma_{\rm{LL}}$ is the correlation length of the effective potential.
When $\xi < 1$, the kinetic energy of the gas becomes negligible and the modulations of the original potential govern the spatial behaviour of the wavefunction $\psi_{0}$, as noticed at the beginning of this Section.
Under this condition, the most appropriate description is the one provided by the usual Thomas-Fermi approximation~\cite{SanchezPalencia:Disorder1DGPEThomasFermi:2006}:
\begin{equation}
\label{eqn:2084}
|\psi_{0}^{\mathrm{TF}}(x)| = \begin{dcases}\sqrt{\frac{\mu-V(x)}{g(N-1)}}& \mu\geq V(x)\\
0 & \mu < V(x)\\
\end{dcases}\hspace{0.05cm}\text{.}
\end{equation}

By considering the same disorder configuration as in Fig.~\ref{fig:EffPotWF_1D_Num_Gvar}a, we compute the ground state of the GPE for three values of the nonlinear coefficient: $g(N-1)=2,020, 101, 5.04$. 
The SP states lying at energies equal to the chemical potentials associated to the interacting states are localized for $g(N-1)=5.04$ and $g(N-1)=101$, and extended for $g(N-1)=2,020$ in the random potential displayed in Fig.~\ref{fig:EffPotWF_1D_Num_Gvar}a, due to the finiteness of the domain. In particular, for $g(N-1)=5.04$ (resp. $g(N-1)=101$), the localization length amounts to $\lambda=12$ (resp. $\lambda=123$), which is larger than the healing length $\xi=2.47$ (resp. $\xi=1.48$). 

The estimate of $\psi_{0}$ based on Eq.~\eqref{eqn:0083} is shown in Fig.~\ref{fig:WaveFx_HighNg_Rep_Approxes}, where the GS is compared to the usual Thomas-Fermi approximation in Eq.~\eqref{eqn:2084} and to the smoothing approximation in Eq.~\eqref{eqn:0070} in Figs.~\ref{fig:WaveFx_HighNg_Rep_Approxes}a-d, and only to the latter scheme in Figs.~\ref{fig:WaveFx_HighNg_Rep_Approxes}e-f. In Figs.~\ref{fig:WaveFx_HighNg_Rep_Approxes}a-b, we show that, for strong interactions (as $g(N-1)=2,020$), when also both the conditions $\xi < 1$ and $V_{0} \ll \mu$ are fulfilled, the Thomas-Fermi approximation proves to be an excellent approximation, since the overlap integral with $\psi_{0}$ reaches $0.9997$. The same value is found also for the ovelap integral between $\psi_{0}$ and $\psi_{0}^{s}$, since in that case $G(x) \approx \delta(x)$, thus $V_{s} \approx V_{\rm{R}}$.
Unlike $\psi_{0}^{\mathrm{TF}}$, $\psi_{0}^{\mathrm{TF,LL}}$ here appears to be far from the GS $\psi_{0}$ of the Gross-Pitaevskii equation, whose behavior is ruled by the original potential $V$ rather than the effective one, $V_{\rm{LL}}$. 

Figs.~\ref{fig:WaveFx_HighNg_Rep_Approxes}c-d represent the case in which $g(N-1)=101$, where the healing length satisfies $\xi \gtrsim 1$ while the chemical potential $\mu \gg V_{0}$. Under these conditions, the Thomas-Fermi approach becomes inadequate, whereas $\psi_{0}^{\mathrm{TF,LL}}$ approaches satisfactorily the wavefunction $\psi_{0}$, with an overlap integral with $\psi_{0}$ which amounts to $0.9994$. At the same time, $\psi_{0}^{s}$ still represents a reliable approximation, since the overlap integral with $\psi_{0}$ is equal to $0.9983$. Figs.~\ref{fig:WaveFx_HighNg_Rep_Approxes}e-f refer to the case in which $g(N-1)=5.04$, characterized by $\mu \sim V_{0}$ and still by $\xi > 1$. Here, the perturbative approach introduced with the smoothing approximation is rather coarse, as well as~--~to a slightly lesser extent~--~the ansatz in Eq.~\eqref{eqn:0083}. 

To summarize, while for strong repulsive interactions, such that $\xi \ll 1$, we have recovered that the stationary state follows the Thomas-Fermi approximation~\cite{SanchezPalencia:Disorder1DGPEThomasFermi:2006}, an analogous scheme based on the effective potential provides an efficient way to compute $\psi_{0}$ for $ 1<\xi < \sigma_{\rm{LL}}$, which holds as long as the gas is not fragmented. It is worthwhile to remark that the delocalizing effect on $\psi_{0}$ in the presence of increasingly strong repulsive interactions also qualitatively agrees with a previous result obtained in the context of the (many-particle) Lieb-Liniger model~\cite{SeiYngZag:2012} with scatterers following the Poisson distribution and Dirichlet boundary conditions on $\psi_{0}$.\\

\subsection{Exploring the mean disorder strength}
\label{subsec:repvardis}
\begin{figure}
\centering
    \includegraphics[width=0.975\columnwidth]{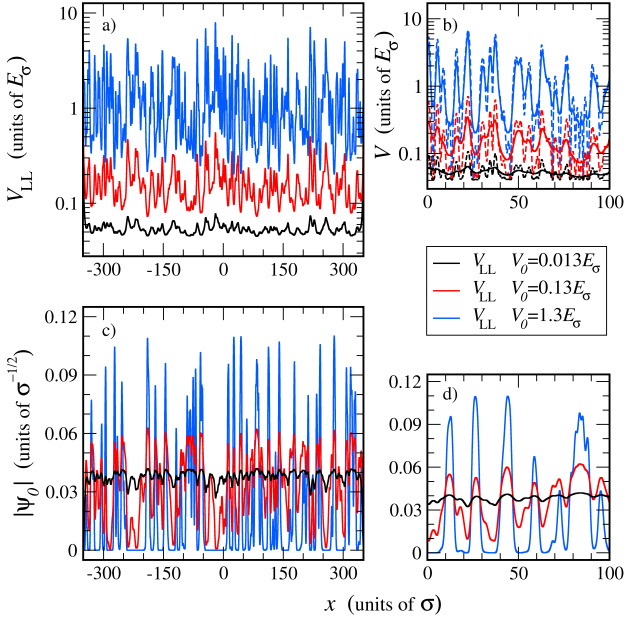}
   \caption{Repulsively-interacting Bose gases with $g(N-1)=30.7$ in speckle potentials with different values of $V_{0}$ and confinement length $l_{\perp}=5.0$ in a domain of length $L=40,000\Delta x$, where $\Delta x=0.0175$. Panel (a): effective potentials $V_{\rm{LL}}$ (solid lines) for different mean values $V_{0}$ of the speckle potential: $V_{0}=0.013$ (black lines), $V_{0}=0.13$ (red lines), $V_{0}=1.3$ (blue lines).
Panel (b): the same quantities as in Panel (a), restricted to the spatial interval $[0,100]$. Original potentials $V$ (dashed lines with the same colors of the corresponding effective potentials).
Panel (c): ground states of the 1D GPE in Eq.~\eqref{eqn:054} for the three mean values $V_{0}$ last mentioned (solid lines).
Panel (d): the same quantities as in Panel (c), but plotted in the same spatial interval as in Panel (b).}
\label{fig:EffPotWF_1D_Num_V0var}
\end{figure}

While in the previous section we have investigated the effect of repulsive interactions on the GS of the GPE, here we asssess the effect of disorder on the spatial behavior of $\psi_{0}$.
Before introducing an efficient scheme to evaluate the particle density in the Lifshitz glass regime, we compute the GS for different values of the parameter $V_{0}$ of the random potential, keeping the correlation length and the transverse-confinement length constant and equal to those set in Ref.~\cite{Billy:AndersonBEC1D:N08}. We begin considering the case of a repulsively interacting gas, using the same parameters as in Fig.~\ref{fig:EffPotWF_1D_Num_Gvar}, except for $g(N-1)=30.7$ and $V_{0}$, which is varied from $0.013$ to $1.3$ by factors of 10. In Fig.~\ref{fig:EffPotWF_1D_Num_V0var}a, we plot the effective potentials $V_{\rm{LL}}$ with solid lines for three different values of $V_{0}$, whereas in Fig.~\ref{fig:EffPotWF_1D_Num_V0var}c we show the corresponding moduli of the wavefunctions $\psi_{0}$.
For the sake of readability, the total potentials $V$ are only plotted in Fig.~\ref{fig:EffPotWF_1D_Num_V0var}b with dashed lines, using the same colors of the corresponding effective potentials. 
As it can be noticed from Fig.~\ref{fig:EffPotWF_1D_Num_V0var}, the modulus of the wavefunction at the maxima of the effective potential $V_{\rm{LL}}$ gets lower as $V_{0}$ is increased from $V_{0}=0.013$ to $V_{0}=1.3$. This reflects the fact that the energy $E_{0}$ becomes progressively smaller than $V_{0}$ as the latter quantity is raised. As a result, the condensate gets more tightly trapped by the potential and ultimately multi-fragmented~\cite{LugCleBou:6:2007,ChengAdhikari:GPEstatsolnum:2010}, i. e., roughly describable as a superposition of localized states.

As this confining effect becomes more pronounced, the effective potential $V_{\rm{LL}}$ gets closer to the original potential, as can be seen in Fig.~\ref{fig:EffPotWF_1D_Num_V0var}b. This effect can be traced back to a \emph{crossover} between the quantum and the semiclassical regimes.
The former regime, occurring for $V_{0}<1$, is characterized by the competition between quantum interference and tunnelling, whereas the latter takes places for $V_{0}>1$ and is dominated by the suppression of tunnelling and the onset of localization due to the barriers of potential. 
Furthermore, in the region of the parameter space explored, the effective potential still allows one to well predict the position of the maxima of the ground state $\psi_{0}$, since $\xi \gtrsim 1$, as seen in Subsec~\ref{subsubsec:strrepint}.
For $V_{0}=0.013$, in the deeply quantum regime,  the SP eigenstate lying at the energy closest to the chemical potential of the state $\psi_{0}$ is also a delocalized state in the domain in Fig.~\ref{fig:EffPotWF_1D_Num_V0var}a, unlike the states associated to the other two GSs, whose healing length is compared to the localization length in Tab.~\ref{tab:Rep_VarDis_Cfrxilmd}.

 \begin{table}[h!]
  \begin{center}
    \begin{tabular}{c|c|c}
      \toprule 
       $V_{0}$ $\left[ E_{\sigma}\right]$  & $\xi$  $\bigl[ \sigma \bigr]$ & $\lambda$  $\bigl[ \sigma \bigr]$\\
       &&\\ [-0.8em]
      \midrule
       0.013 & 2.26 & -\\  
       0.13 & 1.64 & 9.3\\  
       1.3 & 0.956 & 1.2 \\  
      \bottomrule 
    \end{tabular} 
    \caption{Properties of the GSs displayed in Fig.~\ref{fig:EffPotWF_1D_Num_V0var}c-d. Healing length $\xi$ and localization length $\lambda$ of the SP state with energy equal to the chemical potential of the GS of the GPE.}
    \label{tab:Rep_VarDis_Cfrxilmd}
  \end{center}
\end{table}

As noticed in this analysis, for very strong disorder, the ground state $\psi_{0}$ of the Gross-Pitaevskii equation tends to be a superposition of localized SP states which do not exhibit any overlap between each other, unlike those shown in Fig.~\ref{fig:CfrDiagImagTime_87Rb_G01}b.
Since these states belong to the Lifshitz tails of the SP spectrum, the Bose gas becomes a Lifshitz glass, where the gas splits into mini-condensates that occupy the lowest-lying SP states, satisfying the condition $E_{i}^{\mathrm{sp}}\leq \mu$~\cite{LugCleBou:6:2007}.

 According to the landscape theory, these states are expected to occur at the deepest wells of the effective potential $V_{\mathrm{LL}}$, whose occupation number increases as the SP localization length gets larger.
The number of particles $N_{i}^{\mathrm{sp,LL}}$ associated to each one-particle wavefunction $\psi_{i}^{\mathrm{sp,LL}}$ can be thus evaluated, using the LL, as follows:
\begin{equation}
\label{eqn:0754}
N_{i}^{\mathrm{sp,LL}}=\begin{dcases} \frac{\mu^{\rm{LL}}- E_{i}^{\mathrm{sp,LL}}}{U_{i}^{\rm{LL}}} & \text{for} \hspace{0.2cm} \mu^{\rm{LL}} < E_{i}^{\mathrm{sp,LL}}\\
0 & \text{for} \hspace{0.2cm} \mu^{\rm{LL}} \geq E_{i}^{\mathrm{sp,LL}}
\end{dcases}\hspace{0.05cm}\text{,}
\end{equation}
where $N_{s}^{\rm{LL}}$ denotes the number of SP states, $\mu^{\rm{LL}}$ the chemical potential and $U_{i}^{\rm{LL}}:=g\int_{-L/2}^{L/2}\mathrm{d}x\hspace{0.05cm}|\psi_{i}^{\mathrm{sp,LL}}(x)|^{4}$. In the relation last mentioned, the SP state is approximated by means of  Eq.~\eqref{eqn:0075}, while its energy is estimated by exploiting Eq.~\eqref{eqn:0085}.
The numbers of particles in each SP state satisfy 
\begin{equation}
\label{eqn:084}
\sum\limits_{i=0}^{N_{s}^{\rm{LL}}-1}N_{i}^{\mathrm{sp,LL}}=N\hspace{0.05cm}\text{,}
\end{equation}
and the chemical potential associated to the many-particle state can be written as
\begin{equation}
\label{eqn:0753}
\mu^{\rm{LL}} =\frac{N+\sum\limits_{i=0}^{N_{s}^{\rm{LL}}-1}\frac{E_{i}^{\mathrm{sp,LL}}}{U_{i}^{\rm{LL}}}}{\sum\limits_{i=0}^{N_{s}^{\rm{LL}}-1}U_{i}^{\mathrm{LL}\hspace{0.05cm}-1}}\hspace{0.05cm}\text{.}
\end{equation}
Reminding the relation~\eqref{eqn:0072}, the energy of the ground state $\psi_{0}$ of the GPE can be now expressed as:
\begin{equation}
\label{eqn:0755}
E_{0}^{\rm{LL}}=\frac{1}{2N}\sum\limits_{i=0}^{N_{s}^{\rm{LL}}-1}\frac{(\mu^{\mathrm{LL}\hspace{0.05cm}2}-E_{i}^{\mathrm{sp,LL}\hspace{0.05cm}2})}{U_{i}^{\rm{LL}}}\hspace{0.05cm}\text{.}
\end{equation}

\begin{figure}
\centering
    \includegraphics[width=0.975\columnwidth]{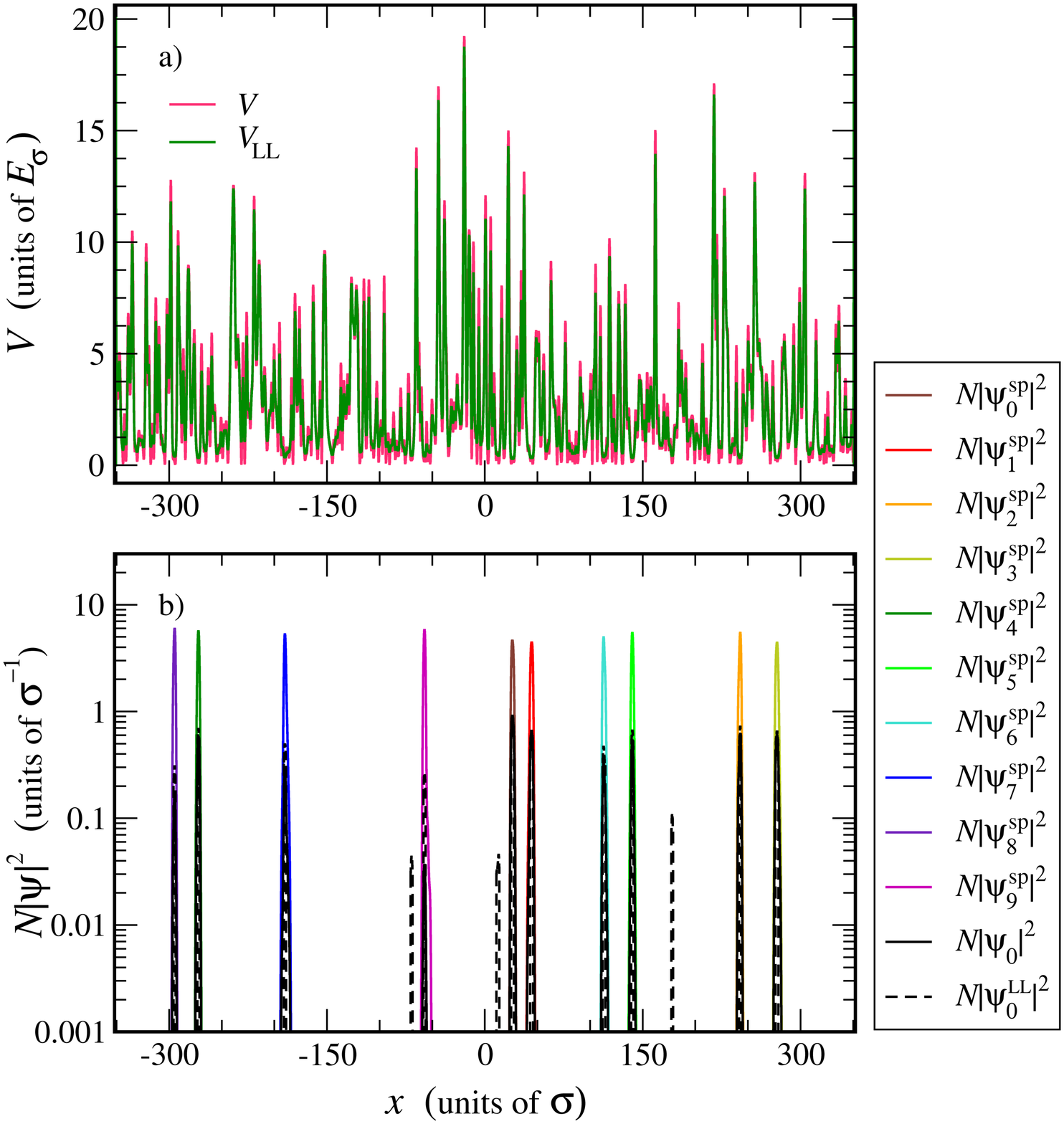}
 \caption{Lifshitz glass phase. A repulsively interacting gas with $g(N-1)=2.71$ and $N=1,500$ atoms of \ce{^{87}Rb} in a 1D speckle potential with typical amplitude $V_{0}=3.0$ in a domain of length $L=40,000\Delta x$ with $\Delta x=0.0175$. Panel (a): original potential $V$ (magenta line) and effective potential $V_{\rm{LL}}$ (green line). Panel (b): the particle densities $\{N|\psi_{i}^{\rm{sp}}|^{2}\}$ associated to the 10 lowest-lying SP eigenstates of $H^{\mathrm{sp}}$ (solid lines), determined by exact diagonalization. Particle density $N|\psi_{0}|^{2}$ related to the ground state of the GPE computed by imaginary-time evolution (black solid line), and the one evaluated by approximating the SP eigenstates using Eq.~\eqref{eqn:0075}, denoted as $N|\psi_{0}^{\rm{LL}}|^{2}$ (black dashed line).}
\label{fig:GPE_LifshitzTailsWF_Rep}
\end{figure}

To this purpose, we consider a realization of the random potential with a mean value $V_{0}$ increased by almost $70$ times compared to the one represented in Fig.~\eqref{fig:CfrDiagImagTime_87Rb_G01}a and a nonlinear coefficient $g(N-1)=2.71$, fixing the number of particles to $N=1,500$.
Despite the high value of $V_{0}$, the spacing between the single-particle energy levels is still lower than $l_{\perp}^{-1}$, thus justifying again the factorization of $\psi_{0}(\pmb{r})$ in Eq.~\eqref{eqn:6574}. 
The ground state computed by imaginary-time evolution, portrayed as the black curve in Fig.~\ref{fig:GPE_LifshitzTailsWF_Rep}b, is nonvanishing along the support of the first $N_{s}=10$ lowest-lying eigenstates of the single-particle problem~\eqref{eqn:3157}, computed by exact diagonalization and represented by the solid lines (from brown to violet). The SP eigenstate $\psi_{9}^{\rm{sp}}$ located at the energy closest to the chemical potential is localized and characterized by a localization length $\lambda=0.57$, which is here smaller than both the correlation length $\sigma$ and the healing length $\xi=0.998$.

Compared to the case examined in Sec.~\ref{subsubsec:wearepint}, $N_{s}^{\rm{LL}}$ here is no longer predicted by Eq.~\eqref{eqn:0087}, which, in the case of Fig.~\ref{fig:GPE_LifshitzTailsWF_Rep}, would lead to an overestimate of that number, thus yielding an upper energy limit for the SP states $E^{\rm{th,LL}}=0.869$ because the interaction is not small enough. 
Since the exact lowest-lying SP states do not occupy the lowest minima of the effective potential in rigorous ascending order of energy, the LL-based predictions concern a slightly different set of SP eigenstates, as shown in the fifth column of Tab.~\ref{tab:TabRepLG-a}.
Hence, both the total energy and the chemical potential in the third column of Tab.~\ref{tab:TabRepLG-b} appear to differ from the estimates above mentioned, by about $5\%$ and $7\%$ respectively.
Besides, the exact diagonalization allows one to determine values of the energy and the chemical potential (in Eqs.~\eqref{eqn:0755} and~\eqref{eqn:0753} respectively) which are in excellent agreement with those associated to the exact numerical GS $\psi_{0}$, as witnessed by the values exposed in the columns 1-2 in Tab.~\ref{tab:TabRepLG-b}.  

The occupation numbers $\{N_{i}^{\mathrm{sp,LL}}\}$ are compared to those evaluated by exact diagonalization, denoted as $\{N_{i}^{\mathrm{sp}}\}$ in Tab.~\ref{tab:Cfr_NumPart_LifshitsStates}a. In the same table, the occupation numbers $\{N_{i}^{\mathrm{sp,GPE}}\}$, extracted by integrating the GS of the GPE along the support of each SP wavefunction, are also reported.
The boundaries of each support are numerically estimated by identifying the points where $\psi_{i}^{\mathrm{sp}}(x)$ decreases to values lower than $0.05$. As it can be inferred, the three estimates agree quite well for the eight lowest-energy states, for which the discrepancy with respect to $\{N_{i}^{\mathrm{sp,GPE}}\}$ amounts to $13\%$ on average for the $\{N_{i}^{\mathrm{sp,LL}}\}$, and to the $1.3\%$ for the $\{N_{i}^{\mathrm{sp}}\}$.
The evaluation of $\{N_{i}^{\mathrm{sp,LL}}\}$ then permits one to obtain the GS $\psi_{0}^{\mathrm{LL}}$, represented as the black dashed curve in Fig.~\ref{fig:GPE_LifshitzTailsWF_Rep}b. As one can also infer from Fig.~\ref{fig:GPE_LifshitzTailsWF_Rep}b, the LL is able to capture all main peaks of the numerical GS $\psi_{0}$, but it also predicts nonzero occupancy of the states $\psi_{10}^{\mathrm{sp}}$, $\psi_{11}^{\mathrm{sp}}$ and $\psi_{12}^{\mathrm{sp}}$. The overlap integral $\int_{-L/2}^{L/2}\psi_{0}^{*}(x)\psi_{0}^{\mathrm{LL}}(x)\hspace{0.05cm}\mathrm{d}x$ between the collective wavefunctions amounts to 0.853. This result suggests a quite good accuracy of the landscape-based scheme in the Lifshitz glass phase.

\begin{table}[h!]
  \begin{center}
    \subfloat[\label{tab:TabRepLG-a}]{
    \centering
    \begin{tabular}{l|c|c|c|c}
      \toprule 
      $i$  & $E_{i}^{\mathrm{sp}}$ $\left[10^{-1}E_{\sigma}\right]$ & $N_{i}^{\mathrm{sp,GPE}}$ & $N_{i}^{\mathrm{sp}}$  & $N_{i}^{\mathrm{sp,LL}}$ \\
      \midrule
      0 & 3.443 & 316 & 310 & 262\\ 
      1 & 3.931 & 230 & 226  & 200\\  
      2 & 4.039 & 180 & 178 & 152\\ 
      3 & 4.040 & 218 & 214 & 194\\ 
      4 & 4.071 & 160 & 160 & 148\\ 
      5 & 4.156 & 154 & 155 & 148 \\ 
      6 & 4.506 & 105 & 107 & 125\\ 
      7 & 4.583 & 87 & 88 & 105 \\ 
      8 & 4.757 & 41 & 44 & 64\\
      9 & 4.957 & 9 & 13 & 55\\
      10 & 5.186 & 0 & 0 & 13\\
      11 & 5.317 & 0 & 0 & 9\\
      12 & 5.425 & 0 & 0 & 19\\
     \bottomrule
     \end{tabular}
     }\hfill
     \subfloat[\label{tab:TabRepLG-b}]{
     \centering
     \begin{tabular}{c|c|c}
     \toprule
     \multicolumn{3}{c}{Ground-state energy} \\ \hline 
      &&\\ [-0.8em]
     $E_{0}$ $\left[E_{\sigma}\right]$ & $E_{0}^{\mathrm{GPE}}$ $\left[E_{\sigma}\right]$ & $E_{0}^{\mathrm{LL}}$ $\left[E_{\sigma}\right]$\\
     \midrule
     $4.526\cdot 10^{-1}$ & $4.515\cdot 10^{-1}$ & $4.732\cdot 10^{-1}$\\
     \toprule
     \multicolumn{3}{c}{Chemical potential} \\ \hline
      &&\\ [-0.8em]
     $\mu$ $\left[E_{\sigma}\right]$ & $\mu^{\mathrm{GPE}}$ $\left[E_{\sigma}\right]$ & $\mu^{\mathrm{LL}}$ $\left[E_{\sigma}\right]$\\
     \midrule
     $5.041\cdot 10^{-1}$ & $5.016\cdot 10^{-1}$ & $5.349\cdot 10^{-1}$\\
     \toprule
    \end{tabular}
    }
    \caption{Lifshitz glass phase. Table (a): summary of the ten lowest-lying SP energy values $\{E_{i}^{\rm{sp}}\}$ and the numbers of particles in each SP state (estimated by using three different methods), related to the system in Fig.~\ref{fig:EffPotWF_1D_Num_V0var}. From column 1 to 2: SP state labels and SP state eigen-energies, computed by exact diagonalization.
From column 3 to column 5: the numbers of bosons in each SP state computed by integration around the peaks of the ground state of the GPE ($N_{i}^{\mathrm{sp,GPE}}$), the ones evaluated by using Eq.~\eqref{eqn:0754} with SP states by exact diagonalization ($N_{i}^{\mathrm{sp}}$), the ones estimated by means of Eq.~\eqref{eqn:0754} with SP states by LL ($N_{i}^{\mathrm{sp, LL}}$).
Table (b): the total energy of the Bose gas and its chemical potential, estimated starting from the data found by the three methods above mentioned.}
    \label{tab:Cfr_NumPart_LifshitsStates} 
  \end{center}
\end{table}

The results of this subsection show that, for increasing disorder mean amplitude and fixed repulsive interaction, the delocalized ground state becomes increasingly fragmented. This ultimately gives rise to the appearance of the Lifshitz glass phase, where the Bose gas splits into mini-condensates localized on SP states which do not spatially overlap and belong to the Lifshitz tails of the SP IDoS. Here the GS of the GPE can be effectively approximated by exploiting the estimates of the SP states obtained by using the LL. We also point out that finiteness of the number of wells occupied by the wavefunction $\psi_{0}$ under those conditions was also proved by Seiringer \textit{et al.}~\cite{SeiYngZag:2012} within the Lieb-Liniger model with randomly distributed potential barriers. 

\section{Conclusions and perspectives}
\label{sec:conclupers}

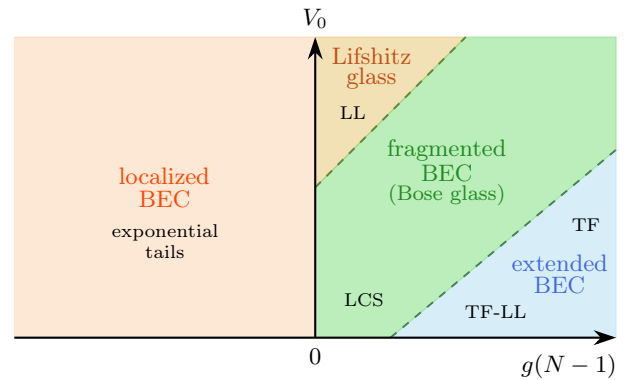
\begin{figure}
\centering 
\begin{tikzpicture}
\draw[line width=0.25mm, dashed, color=darkgray] (4.0,2.0)--(6.0,4.0);
\draw[line width=0.25mm, dashed, color=darkgray] (5.0,0.0)--(8.0,2.5);
\draw[opacity=0.333,fill=SkyBlue,color=SkyBlue] (5.0,0.0)--(8.0,0.0)--(8.0,2.5);   
\draw[opacity=0.333,fill=Goldenrod,color=Goldenrod] (4.0,4.0)--(6.0,4.0)--(4.0,2.0);
\draw[opacity=0.333,fill=LimeGreen,color=LimeGreen] (4.0,0.0)--(5.0,0.0)--(8.0,2.5)--(8.0,4.0)--(6.0,4.0)--(4.0,2.0);
\draw[opacity=0.333,fill=Apricot,color=Apricot] (0.0,0.0)--(4.0,0.0)--(4.0,4.0)--(0.0,4.0);     
\draw[line width=2.0mm, thick, ->,-{Stealth[length=3mm, width=2mm]}] (0.0,0.0) -- (8.0,0.0);
\draw[line width=2.0mm, thick, ->,-{Stealth[length=3mm, width=2mm]}] (4.0,0.0) -- (4.0,4.0);
\node at (4.0,-0.25) {\small{0}};
\node at (6.4,0.35) {\textcolor{black}{\scriptsize{TF-LL}}}; 
\node at (7.6,1.5) {\textcolor{black}{\scriptsize{TF}}}; 
\node at (2.0,1.4) {\textcolor{black}{\scriptsize{exponential}}};
\node at (2.0,1.15) {\textcolor{black}{\scriptsize{tails}}};
\node at (7.25,1.0) {\textcolor{RoyalBlue}{\small{extended}}};
\node at (7.25,0.7) {\textcolor{RoyalBlue}{\small{BEC}}};
\node at (4.75,3.75) {\textcolor{Bittersweet}{\small{Lifshitz}}};
\node at (4.75,3.45) {\textcolor{Bittersweet}{\small{glass}}};
\node at (5.75,2.5) {\textcolor{ForestGreen}{\small{fragmented}}};
\node at (5.75,2.2) {\textcolor{ForestGreen}{\small{BEC}}};
\node at (5.75,1.9) {\textcolor{ForestGreen}{\footnotesize{(Bose glass)}}};
\node at (4.5,3.0) {\textcolor{black}{\scriptsize{LL}}}; 
\node at (4.65,0.5) {\textcolor{black}{\scriptsize{LCS}}}; 
\node at (2.0,2.15) {\textcolor{OrangeRed}{\small{localized}}};
\node at (2.0,1.85) {\textcolor{OrangeRed}{\small{BEC}}};
\node at (7.375,-0.375) {\small{$g(N-1)$}};
\node at (4.0,4.25) {\small{$V_{0}$}};
\end{tikzpicture}
 \caption{Schematic quantum-state diagram in the interaction-disorder plane in which the abreviations denote the approximations used for $\psi_{0}$ throughout the paper: the linear combination of states (LCS) for weakly repulsive interactions, the LL-based approximation in the Lifshitz glass phase, the Thomas-Fermi-like approximation (TF-LL) for intermediate repulsive interactions ($\sigma_{\rm{LL}}<\xi<1$), and the Thomas-Fermi approach for strongly repulsive interactions ($\xi \gg 1$). The dashed lines refer to the crossovers between each phase.}
    
\label{fig:Disorder-Interactions_PhasesMethods}
\end{figure}

We have presented here the numerical computation of the GS $\psi_{0}$ of the quasi-1D Gross-Pitaevskii equation with Gaussian-correlated speckle potentials for a broad range of disorder parameters and nonlinear couplings. The spatial behavior of $\psi_{0}$ has been analyzed in relation with the original potential $V$ and the effective potential $V_{\rm{LL}}$, which is given by the reciprocal of the LL function.
New approaches, based on the LL have been introduced for $\psi_{0}$ in the different phases (shaded areas in the quantum-state diagram in Fig.~\ref{fig:Disorder-Interactions_PhasesMethods}), in the regions of the interaction-disorder plane pinpointed by the violet signs.

For attractive interactions, which strengthen the exponential localization of the atoms, we have shown that the LL allows us to accurately predict the localization center of the GS of the GPE. The behavior of the localization length as a function of both the nonlinear coefficient and the disorder parameter would be worth investigating beyond the weak-disorder regime~\cite{Mueller:GPE:2011,Delande:LocalizationAttrSolitons:2013}, in which the Born approximation holds.

For weak repulsive interactions $g(N-1) \lesssim 1$, we have proved that the ground state of the Gross-Pitaevskii equation is well approximated by an expansion in terms of a finite number of single-particle states $\{\psi_{i}^{\mathrm{sp}}\}$. For intermediate repulsive interactions, when $1<\xi < \sigma_{LL}$, we have assessed an approximation of $\psi_{0}$, based on a Thomas-Fermi-like ansatz using the effective potential, showing that, in this regime, $\psi_{0}$ follows the modulations of the effective potential rather than those of the original one. Nevertheless, when this approximation breaks down, the expansion in SP states is not convenient anymore, since $g(N-1)$ can be much larger than $1$ and a suitable theoretical approach is still missing.

The LL has also opened the possibility to reckon the occupation number of each SP state when the Bose gas is in the Lifshitz glass regime. In order to pinpoint the crossover region between this regime and the Bose glass one, it would be of particular prominence to compute the disorder-averaged atomic population and domain size of the SP states as a function of the interaction strength and the mean value of the random potential, starting from the effective potential.

The approaches here introduced have managed to increase our knowledge of the solutions of the time-independent GPE. These schemes can also be applied for positive-valued random potentials endowed with any spatial distribution, on the condition that the correlation profile has a finite range. For long-range correlations in the disorder, the above methods might be not suitable since they can hinder localization, leading to the occurence of mobility edges even in the noninteracting case~\cite{IzraKrok:1999} and they inhibit the fragmentation occurring for defocusing GPEs~\cite{FalNatPok:2009}.

Furthermore, the results presented in this work can help to structure systems with higher-dimensional random potentials, whose phase diagram is more partially known~\cite{Bourdel:2012,AstKruNav:2013, CarBoeHol:4:2013} and where the features of the wavefunctions have been estimated only for superpositions of Gaussian and harmonic potentials~\cite{FalNatPok:2009}, between the latter and the speckle ones~\cite{SanchezPalencia:Disorder1DGPEThomasFermi:2006}, as well as for Bernoulli potentials~\cite{StaMasBis:6:2012}. For instance, the application of the LL theory to the density of states~\cite{DesforgesFiloche:IDOS:2020} may be helpful in accounting for the numerical phase boundary between the normal and the superfluid phase~\cite{Bourdel:2012, CarBoeHol:4:2013} which occurs in 2D random potentials.

Moreover, it would be of interest to understand how our results would be modified in the presence of corrections to the GPE, like those accounting for beyond-mean-field effects such as quantum fluctuations~\cite{Salasnich:2018}, and those due to finite-range interactions~\cite{Collin:BeyondMFGPE:2007,Cappellaro:BeyondMFGPEDis:2020}, modeling van der Waals potentials.
Another possible application of our methods can be seeked in the Bogoliubov excitations on top of the GS, whose dispersion relation has been found also in random potentials~\cite{Gaul:BogoliubovExcitationsOfDisorderedBEC:PRA11} in arbitrary dimensions.

\begin{acknowledgments}
We gratefully acknowledge V. Josse, T. Bourdel, A. Aspect, D. N. Arnold, S. Mayboroda, N. Cherroret, D. Delande, J.-P. Banon, R. Da Silva Souza, J. de Dios Pont, L.-A. Razo López, P. Pelletier, A. Seye, M. Kakoi and L. Chen for fruitful discussions.
This work was supported by grants from the Simons Foundation (Grants No. 601944 and 1027116, MF; No. 601950, YM) within the framework of the project \textsl{Localization of Waves}. For the numerical simulations, access was granted to the computational resources of the Mésocentre de Moulon in Gif-sur-Yvette (France). 
\end{acknowledgments}

\renewcommand{\thesection}{\Alph{section}}
\appendix

\renewcommand{\theequation}{A.\arabic{equation}}
\setcounter{equation}{0}

\section*{Appendix A: Evaluation of the coefficients of the expansion for weak repulsive interactions} 
\label{app:coeffeval}

In this subsection we outline the analytical approximation followed to efficiently evaluate the coefficients $\{c_{i}\}$ of the expansion in Eq.~\eqref{eqn:0086} for $N_{s}$ SP states.
By plugging Eq.~\eqref{eqn:0086} into the Gross-Pitaevskii equation~\eqref{eqn:054} and multiplying both members on the left by the eigenstate $\psi_{m}^{\mathrm{sp}}$ of $H^{\mathrm{sp}}$, the following relation among the coefficients $\{c_{i}\}$ is found:
\begin{equation}
\label{eqn:0185}
c_{m}(E_{m}^{\mathrm{sp}}-\mu)=g(N-1)\sum\limits_{j,k,n=0}^{N_{s}-1}c_{j}^{*}c_{k}c_{n}\mathcal{I}_{mjkn}\hspace{0.05cm}\text{,}
\end{equation}
where the sum of the squared moduli of the coefficients is normalized to unity and 
\begin{equation}
\label{eqn:1186}
\mathcal{I}_{mjkn}:=\int\limits_{-L/2}^{-L/2}\psi_{m}^{\mathrm{sp}\hspace{0.025cm}*}(x)\psi_{j}^{\mathrm{sp}\hspace{0.025cm}*}(x)\psi_{k}^{\mathrm{sp}}(x)\psi_{n}^{\mathrm{sp}}(x)\hspace{0.05cm}\mathrm{d}x\hspace{0.05cm}\text{.}
\end{equation}
Reminding the definition of the chemical potential in Eq.~\eqref{eqn:0072}, that equation can be rewritten as:
\begin{multline}
\label{eqn:0186}
c_{m}\biggl(E_{m}^{\mathrm{sp}}-\sum\limits_{j=0}^{N_{s}-1}|c_{j}|^{2}E_{j}^{\mathrm{sp}}\biggr)=\\=g(N-1)\sum\limits_{i,j,k,n=0}^{N_{s}-1}(\delta_{im}-c_{i}^{*})c_{j}^{*}c_{k}c_{n}\mathcal{I}_{ijkn}\hspace{0.05cm}\text{,}
\end{multline}
where $\delta_{im}$ is the Kronecker's delta between the SP eigenstates $\psi_{i}^{\mathrm{sp}}$ and $\psi_{m}^{\mathrm{sp}}$.
By finding the equation~\eqref{eqn:0186} for each coefficient $c_{i}$ with $i=0,1,\dots,N_{s}-1$, one obtains a system of nonlinear coupled equations.
Among the $N_{s}^{4}$ overlap integrals involved in Eq. \eqref{eqn:0186}, the most important contributions are those of the $N_{s}^{3}$ integrals containing at least two identical wavefunctions. 
Moreover, assuming real-valued eigenfunctions, the number of relevant integrals reduces to $N_{s}+\frac{3}{2}N_{s}!\bigl(\frac{1}{N_{s}-2!}+\frac{1}{N_{s}-3!}\bigr)$ and the summations on the right-hand side of Eq. \eqref{eqn:0186} become:

\begin{widetext}
\begin{equation}
\begin{split}
 \sum\limits_{i,j,k,n=0}^{N_{s}-1}(\delta_{im}-c_{i}^{*})c_{j}^{*}c_{k}c_{n}\mathcal{I}_{ijkn} &\approx \sum\limits_{i=0}^{N_{s}-1}(c_{i}^{4}-\delta_{im})c_{m}^{3}\mathcal{I}_{iiii} + \sum\limits_{i \neq j}^{N_{s}-1}\bigl(4c_{i}^{3}c_{j}-c_{i}^{3}\delta_{jm}-3c_{j}c_{m}^{2}\delta_{im}\bigr)\mathcal{I}_{iiij} +\\&+ 3\sum\limits_{i > j}^{N_{s}-1}\bigl(2c_{i}^{2}c_{j}^{2}-c_{i}^{2}c_{m}\delta_{jm}\bigr)\mathcal{I}_{iijj}+6\sum\limits_{\substack{i > j\\ j>k}}^{N_{s}-1}\bigl(2c_{i}^{2}c_{j}c_{k}-\delta_{im}\bigr)\mathcal{I}_{iijk}+\\&+3\sum\limits_{\substack{i > j\\ j>k}}^{N_{s}-1}\bigl(4c_{i}c_{j}^{2}c_{k}-c_{j}^{2}c_{k}\delta_{im}\bigr)\mathcal{I}_{ijjk}+3\sum\limits_{\substack{i > j\\ j>k}}^{N_{s}-1}\bigl(4c_{i}c_{j}c_{k}^{2}-c_{j}c_{k}^{2}\delta_{im}\bigr)\mathcal{I}_{ijkk}\hspace{0.05cm} \text{.}
 \end{split}
 \label{eqn:0187}
\end{equation}
\end{widetext}
Under the same assumptions, the following contributions to the left-hand side of Eq. \eqref{eqn:0187} are therefore neglected:
\begin{equation}
\label{eqn:0188}
6\sum\limits_{\substack{i > j\\ j>k\\k>l}}^{N_{s}-1}\bigl(4c_{i}c_{j}c_{k}c_{l}-c_{j}c_{k}c_{l}\delta_{im}\bigr)\mathcal{I}_{ijkl}\hspace{0.05cm}\text{,}
\end{equation}
since they involve integrals over four different one-particle eigenfunctions. As Eqs. \eqref{eqn:0187} for $m=0,1,...,N_{s}$ are polynomial relations up to the fourth order in the coefficients, manifold solutions are possible, all occurring in couples with opposite signs and satisfying the normalization condition, except for the trivial (and unphysical) one, characterized by all vanishing $\{c_{i}\}$. The most appropriate solution for the ground state of the GPE~\eqref{eqn:054} is then identified as the one which minimizes the total energy $E_{0}$.

\section*{Appendix B: Justification of the approximation of the single-particle states based on the localization landscape}
\label{app:loclandSPstates}

\renewcommand{\theequation}{B.\arabic{equation}}
\setcounter{equation}{0}

In this subsection we provide a mathematical explanation in support of Eq.~\eqref{eqn:0075}, which has been exploited in Secs. and~\ref{subsec:repvardis}. Considering the linear decomposition of the localization landscape on the orthonormal basis  $\{\psi_{i}^{\rm{sp}}\}$ of SP eigenstates, in Dirac's notation it reads:
\begin{equation}
\label{eqn:230}
|u\rangle=\sum\limits_{i}u_{i}|\psi_{i}^{\mathrm{sp}}\rangle\hspace{0.05cm}\text{,}
\end{equation}
where $u_{i}:=\langle u |\psi_{i}^{\mathrm{sp}}\rangle$.
Using the same notation, the definition of the localization landscape in Eq.~\eqref{eqn:902} becomes 
\begin{equation}
\label{eqn:231}
H^{\mathrm{sp}}|u\rangle = |1\rangle\hspace{0.05cm}\text{.} 
\end{equation}
By plugging the right-hand side of Eq.~\eqref{eqn:230} into the left-hand side of Eq.~\eqref{eqn:231} and multiplying both sides by $\langle x |$ on the left, one finds:
\begin{equation}
\label{eqn:232}
\sum\limits_{i} \langle x |H^{\mathrm{sp}}|\psi_{i}^{\mathrm{sp}}\rangle u_{i} = \sum\limits_{i}E_{i}^{\mathrm{sp}}\psi_{i}^{\mathrm{sp}}(x)u_{i}=\langle x | 1\rangle\hspace{0.05cm}\text{,}
\end{equation}
where we have used Eq.~\eqref{eqn:3057} as well as the definition $\psi_{i}^{\mathrm{sp}}(x):=\langle x|\psi_{i}^{\mathrm{sp}}\rangle$. 
By multiplying both sides of Eq.~\eqref{eqn:232} by $\psi_{j}^{\mathrm{sp}\hspace{0.025cm}*}(x)$, and performing integrations over the coordinate $x$ one obtains:
\begin{equation}
\label{eqn:233}
\sum_{i}\int \psi_{j}^{\mathrm{sp}\hspace{0.025cm}*}(x)\psi_{i}^{\mathrm{sp}}(x) E_{i}^{\mathrm{sp}}u_{i}  \hspace{0.05cm} \mathrm{d}x=\int\psi_{j}^{\mathrm{sp}\hspace{0.025cm}*}(x)\hspace{0.05cm} \mathrm{d}x=\langle \psi_{j}^{\rm{sp}}|1\rangle\hspace{0.05cm}\text{.}
\end{equation}
Since $\int \psi_{j}^{\mathrm{sp}\hspace{0.025cm}*}(x)\psi_{i}^{\mathrm{sp}}(x)\hspace{0.05cm} \mathrm{d}x=\delta_{ij}$, $\int |x\rangle \langle x|\hspace{0.05cm} \mathrm{d}x= \mathbbm{1}$, and the speckle potential is always positive-valued (see Eq.~\eqref{eqn:055}), one can deduce the following relation from Eq.~\eqref{eqn:233}:
\begin{equation}
\label{eqn:234}
u_{j}=\frac{\langle\psi_{j}^{\mathrm{sp}}|1\rangle}{E_{j}^{\mathrm{sp}}}\hspace{0.05cm}\text{.}
\end{equation}
By plugging the right-hand side of Eq.~\eqref{eqn:234} into the right-hand side of Eq.~\eqref{eqn:230} and multiplying by $\langle x|$ on the left, the decomposition becomes:
\begin{equation}
\label{eqn:235}
u(x)=\sum\limits_{i}\frac{\langle\psi_{i}^{\mathrm{sp}}|1\rangle}{E_{i}^{\mathrm{sp}}}\psi_{i}(x)\hspace{0.05cm}\text{.}
\end{equation}
Within the effective domain $\Omega_{i}$ associated to the eigenstate $\psi_{i}^{\mathrm{sp}}$ one thus finds that the leading contribution to the localization landscape is the one associated to the $i$-th lowest-lying SP state. As a result one can approximate the state last mentioned as the localization landscape in $\Omega_{i}$, multiplied by a dimensional coefficient, as in Eq.~\eqref{eqn:0075}, which is determined by imposing the normalization condition to the SP eigenstate.

\bibliography{BiblioGPELL}

\end{document}